\documentclass[a4paper,fleqn,usenatbib]{mnras}

\usepackage[T1]{fontenc}
\usepackage{ae,aecompl}

\usepackage{graphicx}	% Including figure files
\usepackage{amsmath}	% Advanced maths commands
\usepackage{amssymb}	% Extra maths symbols

\title[The NGC 454 system: anatomy of a mixed on-going merger]{The NGC 454 system: anatomy of a mixed on-going merger\thanks{Based on observations obtained at the Southern Astrophysical Research (SOAR) 
telescope, which is a joint project of the Minist\'erio da Ci\^encia, Tecnologia, e Inova\c c\~ao (MCTI) da Rep\'ublica Federativa do Brasil, 
the U.S. National Optical Astronomy Observatory (NOAO), the University of North Carolina at Chapel Hill (UNC), 
and Michigan State University (MSU).}}

\author[H. Plana et al.]{
H. Plana$^{1},$\thanks{E-mail: plana@uesc.br}
R. Rampazzo$^{2}$,
P. Mazzei$^{2,}$,
A. Marino$^{2,}$,
Ph. Amram$^{3}$,
A.L.B. Ribeiro$^{1}$
\\
$^{1}$Laborat\'orio de Astrof\'isica Te\'orica e Observational,
Universidade Estadual de Santa Cruz --
45650-000 Ilh\'eus - Bahia Brazil \\
$^{2}$INAF-Osservatorio Astronomico di Padova,
Vicolo dell'Osservatorio 5,
35122 Padova Italy\\
$^{3}$Aix Marseille Univ.,  CNRS, Laboratoire d'Astrophysique de Marseille (LAM)
38 rue Fr\'ed\'eric Joliot-Curie
F-13388 Marseille Cedex 13
France
}

\date{Accepted XXX. Received YYY; in original form ZZZ}

\pubyear{2017}

\begin{document}
\label{firstpage}
\pagerange{\pageref{firstpage}--\pageref{lastpage}}
\maketitle

% Abstract of the paper
\begin{abstract}
This paper focuses on NGC~454, a nearby interacting pair of galaxies (AM0112-554, RR23), composed of an early-type (NGC~454~E) and a star forming late-type companion (NGC~454~W). We aim at characterizing this wet merger candidate via a multi-$\lambda$ analysis, from near-UV to optical using SWIFT--{\tt UVOT}, and mapping the H$\alpha$ intensity ($I$) distribution, velocity ($V_r$), and velocity dispersion ($\sigma$) fields with SAM+Perot-Fabry@SOAR observations. Luminosity profiles suggest that NGC 454 E is an S0. Distortions in its outskirts caused by the on-going interaction are visible in both optical and near-UV frames. In NGC~454~W, the NUV-{\tt UVOT} images and the H$\alpha$ show a set of star forming complexes  connected by a faint tail. H$\alpha$ emission is detected along the line connecting NGC~454~E to the NGC~454 main H\,{\textsc{ii}} complex. We investigate the ($I-\sigma$), ($I-V_r$) ($V_r-\sigma$) diagnostic diagrams of the H\,{\textsc{ii}} complexes, most of which can be interpreted in a framework of expanding bubbles. In the main H\,{\textsc{ii}} complex, enclosed in the UV brightest region, the gas velocity dispersion is highly supersonic reaching 60 km~s$^{-1}$. However, H$\alpha$ emission profiles are mostly asymmetric indicating the presence of multiple components with an irregular kinematics. Observations point towards an advanced stage of the encounter. Our SPH simulations with  chemo-photometric implementation suggest that this mixed pair can be understood in terms of a 1:1 gas/halos encounter giving rise to a merger  in  about 0.2 Gyr from the present stage.

\end{abstract}

% Select between one and six entries from the list of approved keywords.
% Don't make up new ones.

%%%%%%%%%%%%%%%%%%%%%%%%%%%%%%%%%%%%%%%%%%%%%%

%%%%%%%%%%%%%%%%% BODY OF PAPER %%%%%%%%%%%%%%%%%%
\section{Introduction}\label{sec:intro}

Interactions  modify the gravitational potential of the involved galaxies and may lead to their merger.  During the interaction the stellar and gas components 
of each galaxy respond differently to the potential variation.  The outcome is directly measurable  in terms of
 morphology, kinematics and, in general, of the physical properties of each galaxy, such as their star formation rate and AGN activity. 
A comprehensive description of the {\it job of interactions} in shaping galaxies and their properties as investigated in last decades of
extragalactic research is widely presented and discussed by \citet[][and references therein]{Struck2011}.

Pairs of galaxies have been used as probes to study interactions. Well-selected samples of pairs and 
catalogues have been  produced \citep[see 
e.g.][]{Karachentsev1972,Peterson79,Rampazzo1995,Soares95,
Barton2000}. Single studies as well as surveys of pair catalogues have been crucial to reveal several 
interaction effects once compared to isolated/unperturbed galaxy samples \citep[see e.g.][Section 
5.3.2]{Rampazzo16}. 
 
Although the vast majority of pair members have similar morphological types,  a first light on the existence of mixed morphology pairs has been shed by the
\citet{Karachentsev1972} Catalog of Isolated Pairs. \citet{Rampazzo1992} estimated that between as much as 10-25\% of the pairs in any complete (non-hierarchical) sample will be of the mixed morphology type. 
At the beginning of 1990s, studies about this kind of pairs were addressed to ascertain possible 
enhancement of the star formation activity, with respect to non interacting samples, via mid and far 
infrared
observations,  at that time often hampered by a low resolution   \citep[see e.g.][] {Xu1991, Surace1993}. Mixed morphology pairs have
been thought as the cleanest systems where to verify possible mass transfer between the gas rich and the gas poor member, typically the
early-type companion. Several candidates of mixed morphology pairs with star formation and AGN 
activity, fueled by gas transfer between components,
have been indicated \citep[see e.g.][]{deMello1995,Rampazzo1995,deMello96,Domingue2003}. The literature reports in general a star formation  enhancement in wet and
mixed pairs \citep[see e. g.][]{LT1978, Combes1994,Barton2000,Barton2003,Smith2007,Knapen2015,Smith2016}.  

The fate of mixed, gravitationally bound pairs is to merge, the available gas  may trigger star formation 
for some time, but it is still unclear what will be the merger product. 
The role of mixed merger has been investigated by
\citet{Lin2008} who suggested  that roughly 36\% of the present day red 
galaxies, typically early-type galaxies, have experienced a mixed merger. 
According to these authors mixed (and wet) mergers  will 
produce red galaxies of intermediate mass, after the quenching of the star formation,
while the more massive part of the red sequence should be generated by stellar mass growth via 
dry-mergers \citep{VanDokkum2005, Faber2007}.

In the context of star formation, the dynamics of the (ionized, neutral and molecular) gas clouds during interaction is a crucial topic.  
H\,{\small \text{I}} bridges as well as clouds larger than 10$^8$ M$_\odot$ are detected in wet interacting/merging pairs  with 20-40 km~s$^{-1}$  velocity dispersion
\citep[see e.g.][]{Elmegreen1993,Irwin1994,Elmegreen1995}. External gas high velocity dispersion is 
possibly linked to an internal high velocity dispersion of the clouds, increasing 
the star formation efficiency. \citet{Combes1994}
suggest that the enhancement of the  star formation in wet interacting galaxies 
may be connected to an increase of the molecular gas that inflows toward 
the center by tidal torque. There are indication that the brightness distribution of
H\,{\small \text{II}} regions in interacting objects differs from unperturbed ones. Bright H\,{\small 
\text{II}} regions can form by gas flows during interaction. They are on the average brighter 
than in isolated galaxies and have a high internal velocity dispersion (15-20 km~s$^{-1}$) 
as reported by \citet{Zaragoza-Cardiel2015}. 
Furthermore, the number of H\,{\small \text{II}}  regions in  interacting 
objects is bigger than in an isolated galaxies with the same absolute magnitude, suggesting 
that interactions do in fact increase the star formation rate.

\begin{figure}
\includegraphics[width=8.5cm]{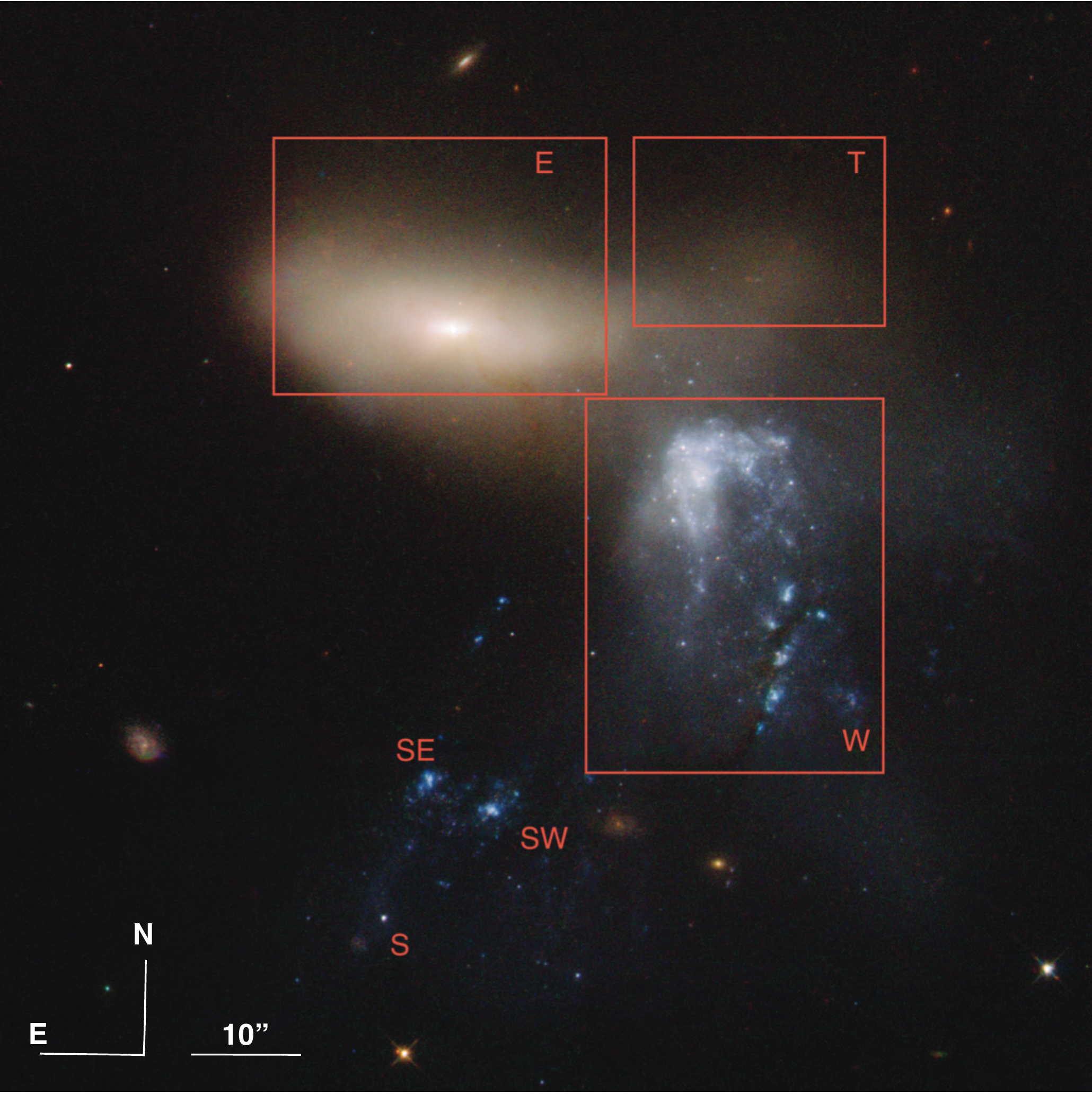} 
\caption{Color composite image of the NGC 454 system obtained with HST-Wide Field Planetary 
Camera 2 in the F450W (B), F606W (V), and F814W (I) filter by  \citet{Stiavelli1998}. 
The figure highlights the different regions of the system following \citet{Johansson1988}  
E and W areas include the early-type and the late-type member of the pair, respectively; 
S, SW and SE are knots likely connected to the W member. T is an area introduced by \citet{Stiavelli1998} (see text). The size of the FOV is 1.66\arcmin.} 
\label{figure1} 
\end{figure}

The subject of the present study is the NGC~454 system, a strongly
peculiar, interacting \citep[AM 0112-554;][]{AM1987} and isolated \citet[RR23;][]{RR95} pair in the Southern Hemisphere. 
\citet{Johansson1988} described the system as ``a pair of emission-line galaxies in close interaction, or in the early-stage of a merger, consisting of a large
elliptical and a blue irregular galaxy". Figure~\ref{figure1} dissects the system according the regions labeled by \citet{Johansson1988} and
\citet{Stiavelli1998}. We will adopt their definition along this paper adding the prefix NGC~454.

The East part of the system, (labeled E in Figure~\ref{figure1}, 
NGC~454~E hereafter),  identifies the early-type member of the pair. NGC~454~E is crossed by dust lanes  and  
it is distorted by the interaction. The U, B, V Johnson and Gunn I photometry by \citet{Johansson1988} presented the East member as a red elliptical with a luminosity profile
that follows closely an r$^{1/4}$ law \citep{deVauc48}  out to 15\arcsec\ from the galaxy center.  The \citet{Stiavelli1998} high resolution HST
imaging, in  the F450W, F606W, and F814W filters, shows that NGC~454~E  is likely an S0. Their luminosity profile,
extending out to $\simeq30\arcsec$, is much better fitted by two components: an  r$^{1/4}$ law  describing the bulge plus an exponential
law \citep{Freeman1970} for a disk.   The (B-V) color profile indicates that the central part of the galaxy, i.e. r$\leq$1\arcsec, is red with
1$\lesssim$ (B-V) $\lesssim$1.4 while the outside region is slightly bluer with 0.8$\lesssim$ (B-V) $\lesssim$1. The nucleus of NGC~454~E,
observed spectroscopically by \citet{Johansson1988}, revealed  several emission lines 
%[OII]$\lambda 3727-29$\AA, [OIII]$\lambda 4363$\AA, $\lambda
%4959$\AA, $\lambda 5007$\AA, [NII]$\lambda 6548$\AA\ and $\lambda
%584$\AA, $H\alpha$$\lambda 6563$\AA\  and [SII]$\lambda 6717$\AA,
%\lambda 6731$\AA. $H\beta$$\lambda 4863$\AA, $H\gamma$$\lambda
%971$\AA, $H\delta$$\lambda 4103$\AA\ are  visible in absorption, while
%$H\epsilon$$\lambda 3971$\AA\ may  fill part of the CaII~H-line $\lambda
%3970$\AA\ less deep than the CaII~K-line $\lambda 3935$\AA.
%\citet{Johansson1988} noticed that the E nucleus 
and matched two of the empirical criteria proposed by \citet{Shuder1981} for a Seyfert galaxy:
the line-width of H$\alpha$ is larger than 300 km~s$^{-1}$ and the [OIII]$\lambda 5007$\AA/H$\beta$ ratio is larger than 3. However, none
of the high-excitation lines expected in this case,  as HeII,  were detected \citep[see also][]{Donzelli2000,Tanvuia2003}. 
The AGN type of  the nucleus  has been  recently detailed by \citet{Marchese2012}. Their
analysis of {\it SWIFT}, XMM-{\it Netwon} and {\it Suzaku} observations characterizes the NGC454~E nucleus as  a ``changing look'' AGN. 
This is a class of AGN  showing significant variation of the  absorbing column density along the line of sight.

The West region of the pair labeled in Figure~\ref{figure1} as W (NGC~454~W hereafter) has been considered by \citet{Johansson1988}  as
the debris of an irregular galaxy. However, the galaxy is so widely distorted by the on-going interaction that the classification is difficult.
 \citet{Stiavelli1998} suggested that it is  the debris of a disk galaxy.  NGC 454 W is a starburst galaxy,  as shown by the  H$\alpha$ image of
\citet[their Figure 6a and 6b]{Johansson1988}.  The spectrum of NGC~454~W shows emission lines whose ratios, 
according to the above authors, are due to photo-ionization by star formation and shock heating.

The NGC~454~E region is particularly distorted in the North-West side. 
This region, label as T (NGC~454~T hereafter) in Figure~\ref{figure1},
has been studied by \citet{Stiavelli1998} which found  this is composed 
by a mix of the stellar populations of the NGC~454~E and the 
NGC~454~W. Moreover, they found a similarity between the color of  
NGC~454~T region and that of the nearby sky and
speculated about the presence of a  faint tail of stripped material 
in this region not detected by the HST observations.

%--------------------------------------- Table 1 ----------------------------------------------
\begin{table}
 \centering
   \caption{NGC 454 system basic properties from the literature}
   \begin{tabular}{@{} llc @{}}
      \multicolumn{3}{c}{} \\
      \hline
 NGC 454 E  &  & Ref.     \\
 Morphology     &  E/S0 pec              &  (1,2) \\
 R.A. (2000)      &   1$^h$ 14$^m$ 25.2$^s$   &    (3)  \\
 Decl. (2000)     &  -55$^\circ$ 23' 47"              &    (3)  \\
%   V$_{hel}$  & 3560$\pm$40 km~s$^{-1}$  & (1) \\
%      V$_{hel}$  & 3587 km~s$^{-1}$  & (3) \\
   V$_{hel}$  & 3635$\pm$2 km~s$^{-1}$  & (3) \\
$ (B-V)_0$& 0.80 &  (1)\\
$(U-B)_0$ & 0.31 & (1) \\   
$L_X$ (0.1-0.3 keV) XMM  &  2.8 10$^{39}$ erg cm$^{-2}$ s$^{-1}$    &  (4) \\
$L_X$ (0.1-0.3 keV) Suzaku  & 5.6 10$^{39}$ erg cm$^{-2}$ s$^{-1}$     &   (4) \\
$L_X$ (0.1-0.3 keV) XMM  &  2.5 10$^{42}$ erg cm$^{-2}$ s$^{-1}$    &  (5) \\
$L_X$ (0.1-0.3 keV) Suzaku  & 7.2 10$^{42}$ erg cm$^{-2}$ s$^{-1}$     &   (5) \\
$L_X$ (14-150 keV) XMM  &  4.8 10$^{42}$ erg cm$^{-2}$ s$^{-1}$  &  (5) \\
$L_X$ (14-150 keV) Suzaku  & 1.4 10$^{42}$ erg cm$^{-2}$ s$^{-1}$     &   (5) \\
   &                     &      \\   
 \hline\hline
  NGC 454 W &   &   \\   
 Morphology     &    Irr, (disrupted) Sp                                        &   (1,2) \\
  R.A. (2000)      &    1$^h$ 14$^m$ 20.1$^s$                   &   (3)   \\
 Decl. (2000)     &      -55$^\circ$ 24' 02"               &     (3) \\
  V$_{hel}$  & 3626$\pm$2 km~s$^{-1}$  & (3) \\
  $(B-V)$& 0.32 & (1)\\
  $(U-B)_0$ & -0.22 & (1) \\   
M(H$_2$) [10$^8$ M$_\odot$]  &         $<2$            &     (6) \\   
      \hline\hline
V$_{hel}$ adopted  [km~s$^{-1}$ & 3645   & (7)\\    
Distance [Mpc] & 48.5$\pm$3.4  &   (7)\\    
scale [kpc arcsec$^{-1}$] & 0.235 &(7) \\
\hline\hline
   \end{tabular}
\label{table1} 

References:   
(1) \citet{Johansson1988} provides the mean corrected radial velocity; 
the morphology is uncertain (2) \citet{Stiavelli1998}; 
(3) the Heliocentric velocities of the E and W components are derived
from \citet{Tanvuia2003} and are consistent with the
systemic velocity V$_{hel}$=3645 provided by {\tt NED} we adopted;
(4) Braito Valentina private communication; (5)  \citet{Marchese2012}
(6) \citet{Horellou1997}; (7) The adopted heliocentric velocity 
and the distance (Galactocentric GSR) of the NGC 454 system are from {\tt NED}.
\end{table}
%--------------------------------- end Table 1 ---------------------------------------------------

The picture of the NGC 454 system  is completed by three blue knots,
NGC~454~SW, NGC~454~SE and NGC~454~S,
well detached from NGC~454~E, and likely
connected to NGC~454~W  \citep{Johansson1988,Stiavelli1998}. 
%Their masses, 3$\times$10$^6$ M$_\odot$, and the age of their stellar
%populations, 1-5$\times$10$^7$\,Gyr, have been estimated by
 \citet{Johansson1988} suggested that these are newly formed
globular clusters  of 3$\times$10$^6$ M$_\odot$ and 1-5$\times$10$^7$\, years  
stellar age. Recently several investigations suggest that 
young independent stellar systems at $z\simeq0$ start to 
form in tidal debris  \citep[see e.g. review by][]{Lelli2015}.\\
\indent
Table~\ref{table1} summarizes some  basic characteristics of
NGC~454~E and W which appear as a prototype of an encounter/merger
($\Delta$V$_{hel}$=1$\pm$2 km~s$^{-1}$ \citet{Tanvuia2003})
between  a late and an early-type galaxy, this latter
having an active and peculiar Seyfert-like nucleus.
Therefore, the study of this system can make progresses in our  
understanding of the effects of a wet interaction.\\
\indent
Our contribution consists of two correlated pieces
of observational information: (1) the investigation of the multi-wavelength
structure of NGC~454~E
via the optical and UV surface photometry and (2) the analysis 
of the kinematics of the ionized gas component. 
The {\it Galaxy Evolution Explorer} \citep[{\it GALEX}]{Morrissay2007} showed the 
 strength of UV observations in revealing rejuvenation episodes
 in otherwise old stellar systems \citep[see e.g.][and references 
 therein]{Rampazzo2007,Marino2011, Rampazzo2011}. 
 In this context, we investigate the Near UV 
 (NUV hereafter) stellar structure of the  NGC~454 system using
 {\it Swift} {\tt UVOT} UV images
 \citep[see also][and references therein]{Rampazzo2017}.
In order to study the ionized gas, connected to the ongoing star formation,
we use the SOAR Adaptive Module (SAM) 
coupled with a Fabry-Perot to investigate its distribution, 
the 2D velocity and velocity dispersion fields of 
the H$\alpha$ emission in NGC~454~W and in 
 NGC~454~SW and SE stellar complexes, likely debris of NGC~454~W.
 We finally attempt to derive the parameters and the merger history
 of the system using simulations.  \\
\indent
The paper is organized as follows. In Section~\ref{sec:OR} we present
{\it SWIFT}-{\tt UVOT}  (\S~\ref{subsec:UVOT}) and 
the SAM+Fabry-Perot (\S~\ref{subsec:FP}) observations of NGC~454 
and the reduction techniques used. The {\it Swift}-{U\tt VOT} surface 
brightness photometry is presented in  \S~\ref{sec:UVOTresults}.
The ionized gas kinematics is presented  in \S~\ref{sec:FPresults}, the diagnostic
digrams of H\,{\textsc{ii}} complexes are discussed in \S~\ref{sec:Diagnostic diagrams},
while   \S~\ref{sec:Profile decomposition} considers the H$\alpha$ line
profile decomposition.  In  \S~\ref{sec:discussion} our results are
discussed in the context of galaxy-galaxy interaction and compared
with Smoothed Particle Hydrodynamic (SPH) simulation with chemo-photometric
implementation in \S~\ref{subsec:simulation}.  Finally in  \S~\ref{sec:summary}, we give the summary and draw general conclusion.

%%%%%%%%%%%%%   FIGURE 1 - HST IMAGE %%%%%%%%%%%%%%%%%%%
 %%%%%%%%%%%%% END FIGURE 1 %%%%%%%%%%%%%%%%%%

\section{Observation and Data reduction} \label{sec:OR}

\subsection{SWIFT-{\tt UVOT} observations} \label{subsec:UVOT}

{\tt UVOT} is a 30~cm telescope in the {\it SWIFT} platform operating
both in imaging and spectroscopy modes \citep{Roming2005}.
We mined the {\tt UVOT } archive in the ASDC-ASI Science Data Center
retrieving the 00035244003 product including images of the NGC 454 system 
in all six filters available. Table~\ref{table3} gives the characteristics of these filters and calibrations are discussed in \citet{Breeveld10,Breeveld11}.
%$UVW2$ ($\lambda_0 \ 2030$), $UVM2$ ($\lambda_0 \ 2231$), 
%$UVW1$ ($\lambda_0 \ 2634$), $U$ ($\lambda_0 \ 3501$),
%%$B$ ($\lambda_0 \ 4329$), $V$ ($\lambda_0 \ 5402$). Description
%of the filters, PSFs FWHM (2\farcs92 for $UVW2$, 2\farcs45 for $UVM2$,
%2\farcs37 for $UVW1$, 2\farcs37 for $U$, 2\farcs19 for $B$, 2\farcs18 for $V$),  and 
%calibrations are discussed in \citet{Breeveld10,Breeveld11}. 

The archival {\tt UVOT} un-binned images  have a scale of 0\farcs5/pixel. Images 
were processed using the procedure described in {\tt http://www.swift.ac.uk/analysis/uvot/}.
All the images taken in the same filter are combined  
in a single image using {\tt UVOTSUM} to improve the S/N and to enhance the visibility 
of NUV features of low surface brightness. The final data set of the 
$UVW2$, $UVM2$, $UVW1$, $U$, $B$, $V$
images have total exposure times reported in Table~\ref{table3}. 
 
%We used the  photometric zero points  provided by \citet{Breeveld11} for converting
%{\tt UVOT} count rates to the AB magnitude system \citep{Oke74}:
%{\it zp$_{UVW2}$} = 19.11$\pm$0.03, {\it zp$_{UVM2}$} = 18.54$\pm$0.03, 
%{\it zp$_{UVW1}$} = 
%18.95$\pm$0.03, {\it zp$_U$} = 19.36$\pm$0.02, {\it zp$_B$} = 18.98$\pm$0.02 and
%{\it zp$_V$} = 17.88$\pm$0.01. 

{\tt UVOT} is a photon counting instrument and, as such, is subject to 
coincidence loss when the throughput is high, whether due to background 
or source counts, which may result in an undercounting of the flux affecting
 the brightness of the source.  
 %Moreover, this produces  a misplacement
%of the source position due to the centroiding algorithm.   
Count rates less than 0.01 counts~s$^{-1}$~pixel$^{-1}$ 
are affected by at most 1\% and count rate less than 0.1  counts~s$^{-1}$~pixel$^{-1}$ 
by at most 12\% due to coincidence loss  \citep[their Figure 6]{Breeveld11}.

Coincidence loss effects can be corrected in the case of point sources
\citep{Poole2008, Breeveld10}. For extended sources a correction process
has been performed for NGC 4449, a Magellanic-type irregular galaxy with
bright star forming regions, by \citet{Karczewski13}.  Even though their
whole field is affected, the authors calculate that the statistical and
systematic uncertainties in their total fluxes amount to $\approx$ 7-9\%
overall,  for the NUV and the optical bands.
 
 We checked, indeed, that in UV filters 
the coincidence losses may involve only few central pixels of the Irr galaxy, i.e., NGC~454~W, never 
exceeding 0.1 count~s$^{-1}$~px$^{-1}$ (in particular, the maximum value of the count rates is  0.043, 0.028, and 
0.047 count~s$^{-1}$~px$^{-1}$  in $UVW2$,  $UVM2$ and 
$UVW1$ filters respectively). Our NUV images are very
slightly affected so we decided do not account for this effect.
Optical images are more
affected. In the NGC~454~W region the effect remains $\le 0.1$
count~s$^{-1}$~px$^{-1}$ in all the bands, in particular it reaches 0.09, , 0.08,  and 0.096 
count~s$^{-1}$~px$^{-1}$ in
the $U$, $B$ and $V$ filters respectively. As far
as NGC~454~E is concerned, in the $U$ filter count rates  are at most 0.084 
count~s$^{-1}$~px$^{-1}$, and reach 0.2
in the $B$ and $V$ bands  in the inner 5\arcsec. So, we
add to the photometric error in Table 3 a further error of 12\% in optical bands 
to account for this effect.

We compared our total magnitudes in Table~\ref{table3} 
 with \citet{Prugniel98}  which reported a total  magnitude 
B=13.32$\pm$0.05 and 13.44$\pm$0.064, respectively for 
the whole NGC~454 system  and for NGC~454~W. Once our measures
are scaled to the Vega system (B=B[AB]+0.139) we have 
B=13.43$\pm$0.15 and B=13.65$\pm$0.12, in very good agreement with previous estimates.

Our ($B-V$) color, integrated within  a
31\arcsec\ aperture and corrected for galactic absorption for
NGC~454~E and NGC~454~W is 0.88$\pm$0.11 and
0.48$\pm$0.06, respectively, to be compared with 0.80 and 0.32 from  \citet{Johansson1988}.

%We conclude that coincidence loss affect our measures  within photometric errors.

\subsection{Fabry-Perot observation} \label{subsec:FP}

Fabry-Perot (FP hereafter) observations\footnote{All Fabry-Perot data (cubes and moment maps) are available at cesam.lam.fr/fabryperot/} have been carried out on Sept
30th 2016 as part of the SAM-FP Early Science run at SOAR 4.1m telescope
at Cerro Pachon (Chile). SAM-FP is a new instrument, available at SOAR,
combining the adaptative optics SAM \citep{TokovininA2010,TokovininB2010}
and a scanning Queensgate ET70 Etalon \citep[][]{MendesdeOliveira2017}.
The SAM module has been conceived to deliver a 0\farcs35 angular
resolution across a 3\arcmin$\times$3\arcmin\ FoV, depending on
atmospheric condition, using Ground Layer Adaptive Optics (GLAO). The SAM instrument detector is a 4K$\times$4K CCD
with a scale image scale of 0\farcs0454 (physical pixel of 15$\mu$) on
the sky \citep{Fraga2013}. The present observations have been binned
over 4$\times$4 pixels resulting in a scale of 0\farcs18/px. The
interferometer used is a ET70 Queensgate scanning FP with an order
of p=609@H$\alpha$. The FP piezos are
driven by a CS100 controller, positioned at the telescope.
Table~\ref{table2} gives the journal of observation with the
characteristic of the etalon we used.
At the center of the Free Spectral
Range we adopt the systemic velocity of 3645 km~s$^{-1}$ provided  
by {\tt NED}  \citep[see also][as more recent and independent sources]{Donzelli2000,Tanvuia2003}.
\\
\indent
Data have been reduced using home made Python macros to handle Multi
Extension Files from SAM and building the data cube, some {\tt IRAF\footnote{{\tt IRAF} is
distributed by the National Optical Astronomy Observatories, which are
operated by the Association of Universities for Research in Astronomy,
Inc., under cooperative agreement with the National Science Foundation.}}
specific tasks and {\tt Adhocw} \footnote{Available at: https://cesam.lam.fr/fabryperot/index/softwares} software procedures have been used to handle 
the cube.
The data reduction procedure has been extensively described by
\citet{Amram1996} and \citet{Epinat2008}. The first step, before the phase correction, is to
perform the standard CCD data reduction by applying bias and flat-field corrections under {\tt IRAF} as well as the cosmic removal using
{\tt L.A.cosmic} procedure \citep{vanDokkum2001}. Linear combinations of dark images have been used to removed CCD patterns in different frames.

In addition to these
canonical operations, it is necessary to check and correct for several
effects, such as, misalignment of data cube frames (due to bad guiding),
sky transparency variation throughout the cube, and seeing variation.
Misalignment variation across the 43 frames of the data cube is less
than half pixel, thus it is negligible. The sky transparency has been
corrected using a star in the FoV. It varies between 81\% to 98\% during
the observation: each frame has been corrected accordingly using one
channel as a reference. The same star is also used to map the  corrected seeing
variation ranging from 0\farcs71 to 0\farcs90. We then applied a 2D
spatial Gaussian smoothing equivalent to the worse estimated corrected seeing
(0\farcs90). Phase map and phase correction have been performed
using the {\tt Adhocw} package and by scanning of the narrow Ne 6599\AA\
line under the same observing conditions. 
The velocities measured are very accurate compared to the systemic
velocity in Table\,\ref{table1}, with an error of a fraction of a
channel width (i.e., $<$ 3 km~s$^{-1}$) over the whole FoV. The signal
measured along the scanning sequence was separated into two parts: (i)
an almost constant level produced by the continuum light in a 15\AA\
passband around H$\alpha$ (continuum map, not presented in this work);
(ii) a varying part produced by the H$\alpha$ line (H$\alpha$ integrated
flux map). The continuum is computed by taking the mean signal outside
the emission line. The H$\alpha$ integrated flux map was obtained by
integrating the monochromatic profile in each pixel. The velocity
sampling was 11.6 km~s$^{-1}$. Strong OH night-sky lines passing through
the filters were subtracted by determining the level of emission away
from our target \citep{Laval1987}. \\
%In order to improve the signal-to-noise (S/N) ratio we used both spectral (rectangular, three channels) and spatial box car smoothing.\\
\indent
The velocity dispersion ($\sigma$ hereafter) is derived from the determination of the FWHM from the determined profile.
%The velocity dispersion ($\sigma$ hereafter) is derived from the
%single Gaussian fit of the H$\alpha$ line profile. %dividing the FWHM by
% the 2.35 factor.
Then the real dispersion velocity is found supposing that different contributions follow a gaussian function. 
%\citep[see e.g.][]{Rozas2000}:

$$\sigma_{real}^2 = \sigma_{obs}^2 - \sigma_{th}^2 - \sigma_{inst}^2$$

where $\sigma_{inst} = 12.82$ km~s$^{-1}$ is the instrument broadening
deduced from the Ne calibration lamp and  $\sigma_{th} = 9.1$
km~s$^{-1}$ is the thermal broadening of the H$\alpha$ line. 

%%%%%%%%%%%%%%%%%%%%%%%%%%%%%%%%%%%%%%% TABLE 2 %%%%%%%%%%%%%%%%%%%%%%%%%%%%%%%%%%%%

\begin{table}
\caption{Instrumental setup}
\label{table2}
\centering
\begin{tabular}{lc}
\hline \hline
Fabry-Perot Parameters & Values\\
\hline
\hline
Telescope & SOAR 4.1m   \\
Date   & Sept. 30$^{th}$2016   \\
Instrument  & SAM-FP $^a$ \\ 
Detector    &   CCD \\
Pixel size (binned) & 0\farcs18/px (0\farcs0454$\times$4)\\
Calibration neon light ($\lambda$) & 6598.95 {\AA}   \\
Resolution ($\lambda/\Delta \lambda$)&  10700  \\
Filter Characteristics &  \\
~~~~~~~~Filter Central wavelength & 6642\AA\ \\
~~~~~~~~Filter Transmission & 80\%@6642\AA\ \\
~~~~~~~~Filter FWHM ($\Delta \lambda$)&  15\AA\ \\
Interferometer Characteristics &  \\
~~~~~~~~Interferometer order at H$\alpha$  & 609 \\
~~~~~~~~Free spectral range at H$\alpha$ (km s$^{-1}$) &  498 \\
~~~~~~~~Number of scanning steps   & 43 \\
~~~~~~~~Sampling steps  & 0.26 \AA ~~ (11.60 km s$^{-1}$) \\
Total Exposure Time                & 1.1h (90s/channel)     \\ 
\hline
\hline
\footnotesize{$^a$ \citet{TokovininA2010,TokovininB2010}}
\end{tabular}
\end{table}

%%%%%%%%%%%%%%%%%%%%%%%%%%%%%%%%%%%%%%% TABLE 2 %%%%%%%%%%%%%%%%%%%%%%%%%%%%%%%%%%%%

%%%%%%%%%%%%%%%%%%%% FIGURE 2 - UV - COLOR PROFILES %%%%%%%%%%%%%%%%%%%%%%%%%%
\begin{figure*}
\includegraphics[width=8cm, angle=0]{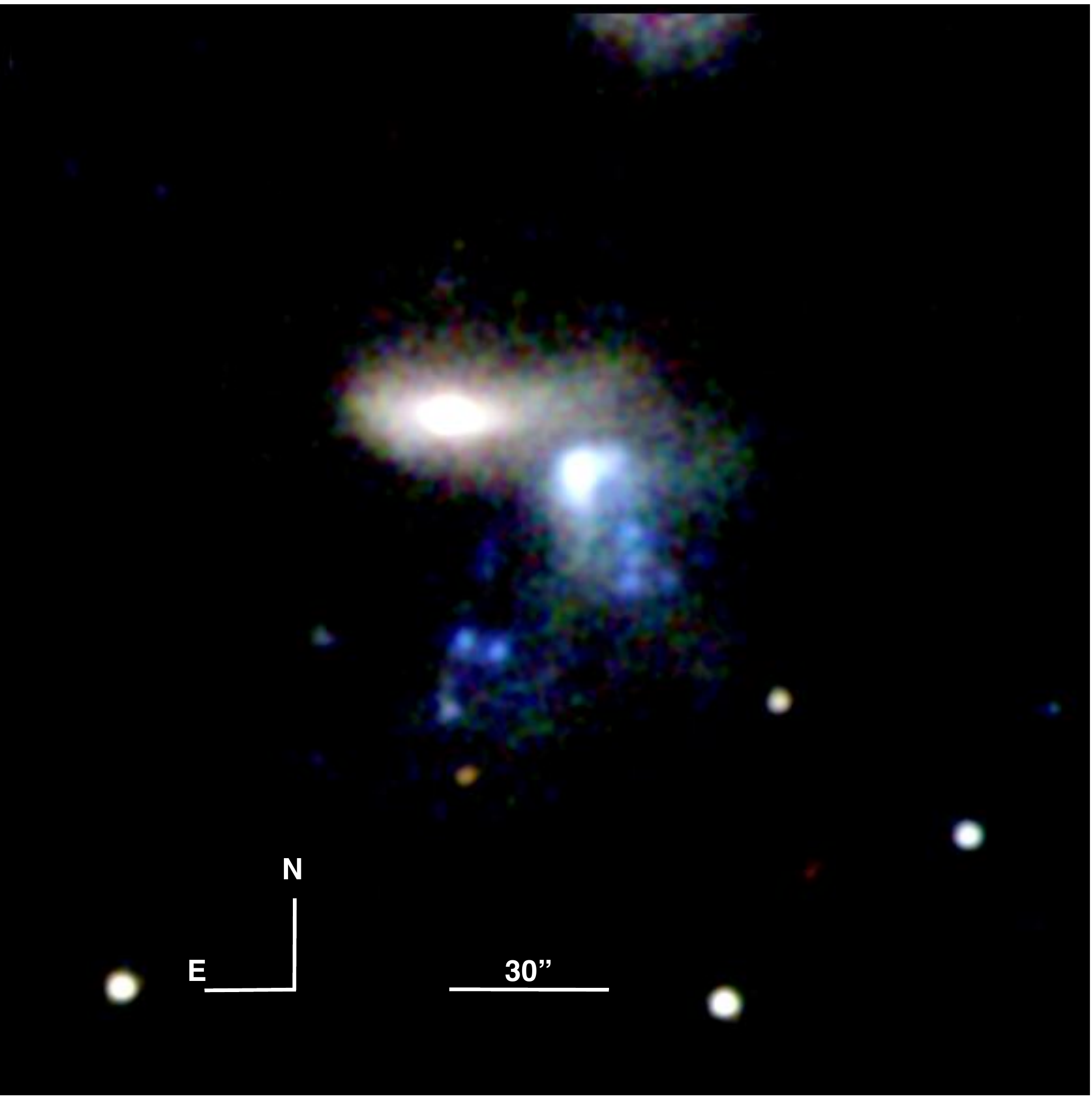}
\includegraphics[width=8cm, angle=0]{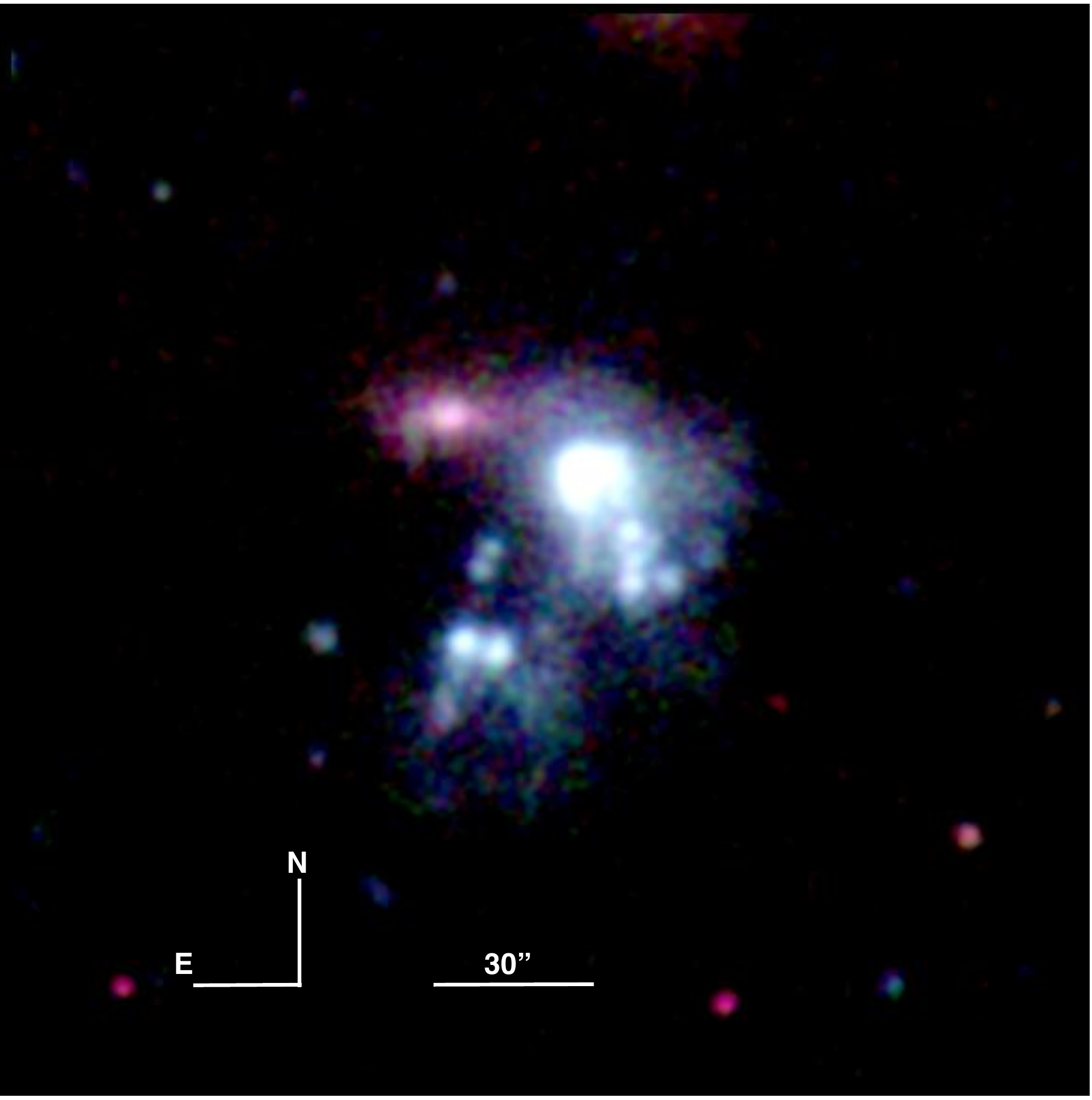}
\includegraphics[width=16cm, angle=0]{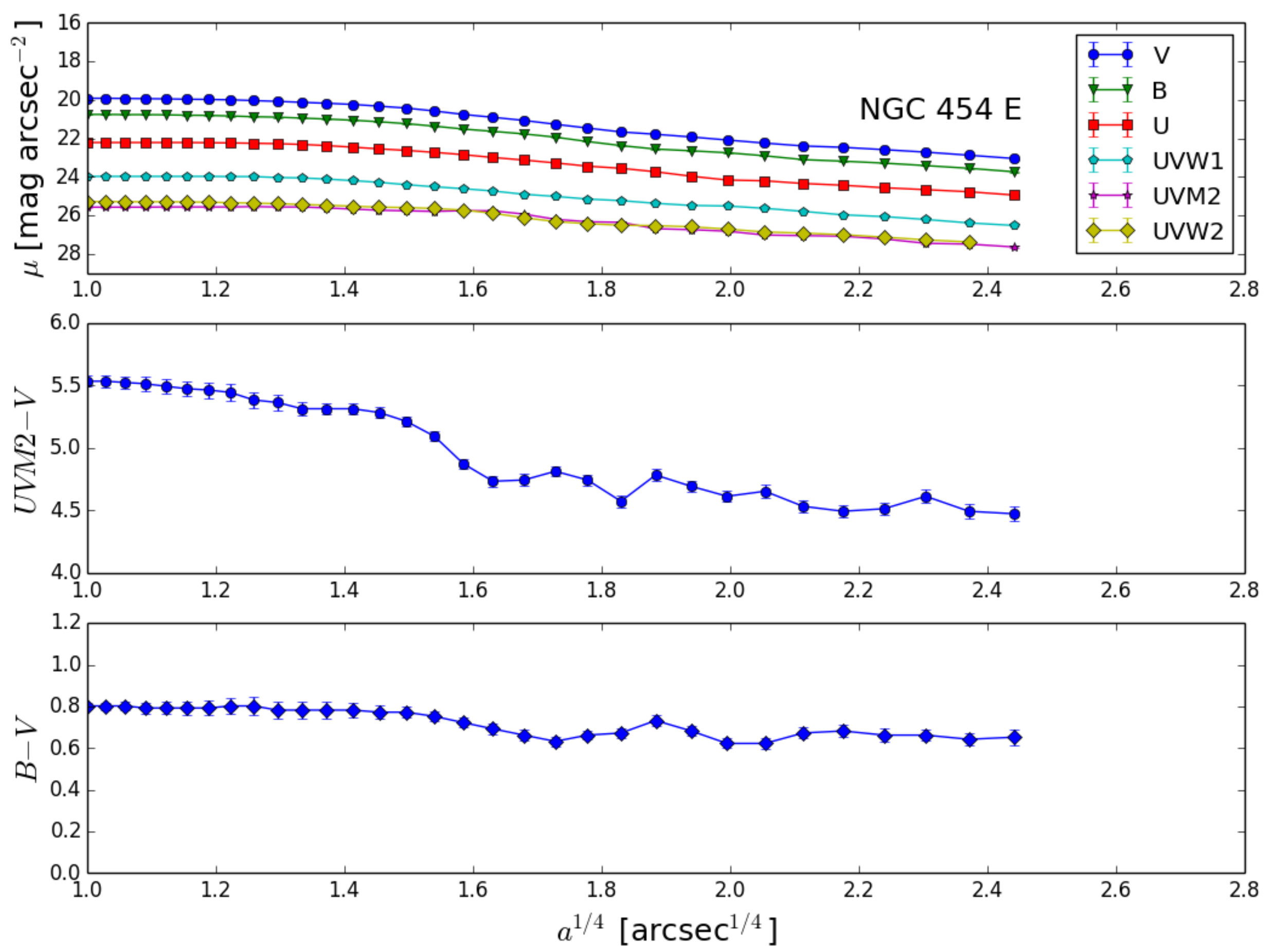}
\caption{(top panels) Optical image ($U~blue$, $B~green$, $V~red$), on the left and UV color composite image ($UVW2~blue$, $UVM2~green $, $UVW1~red$), on the right, of the NGC 454 system as
observed by {\it Swift}-{\tt UVOT}. The images have been smoothed
$2\times2$ pixels (resulting 1\arcsec$\times$1\arcsec resolution). The
total Field of View is $4$\arcmin$\times4$\arcmin. (middle panel) Luminosity profiles
of NGC~454~E in the optical and NUV bands. Profiles are
not corrected for coincidence loss and galactic absorption. (bottom panels) ($M2-V$) and
($B-V$) color profiles in [AB] magnitudes corrected for galactic
absorption.}
\label{figure2}
\end{figure*}
%%%%%%%%%%%%%%%%%%% END FIGURE 2 %%%%%%%%%%%%%%%%%%%%%%%

%%%%%%%%%%%%%%%%%%%% FIGURE 3- \MUOPTICAL - \MU SERSIC%%%%%%%%%%%%%%%%%%%%
\begin{figure}
\includegraphics[width=8.5cm, angle=0]{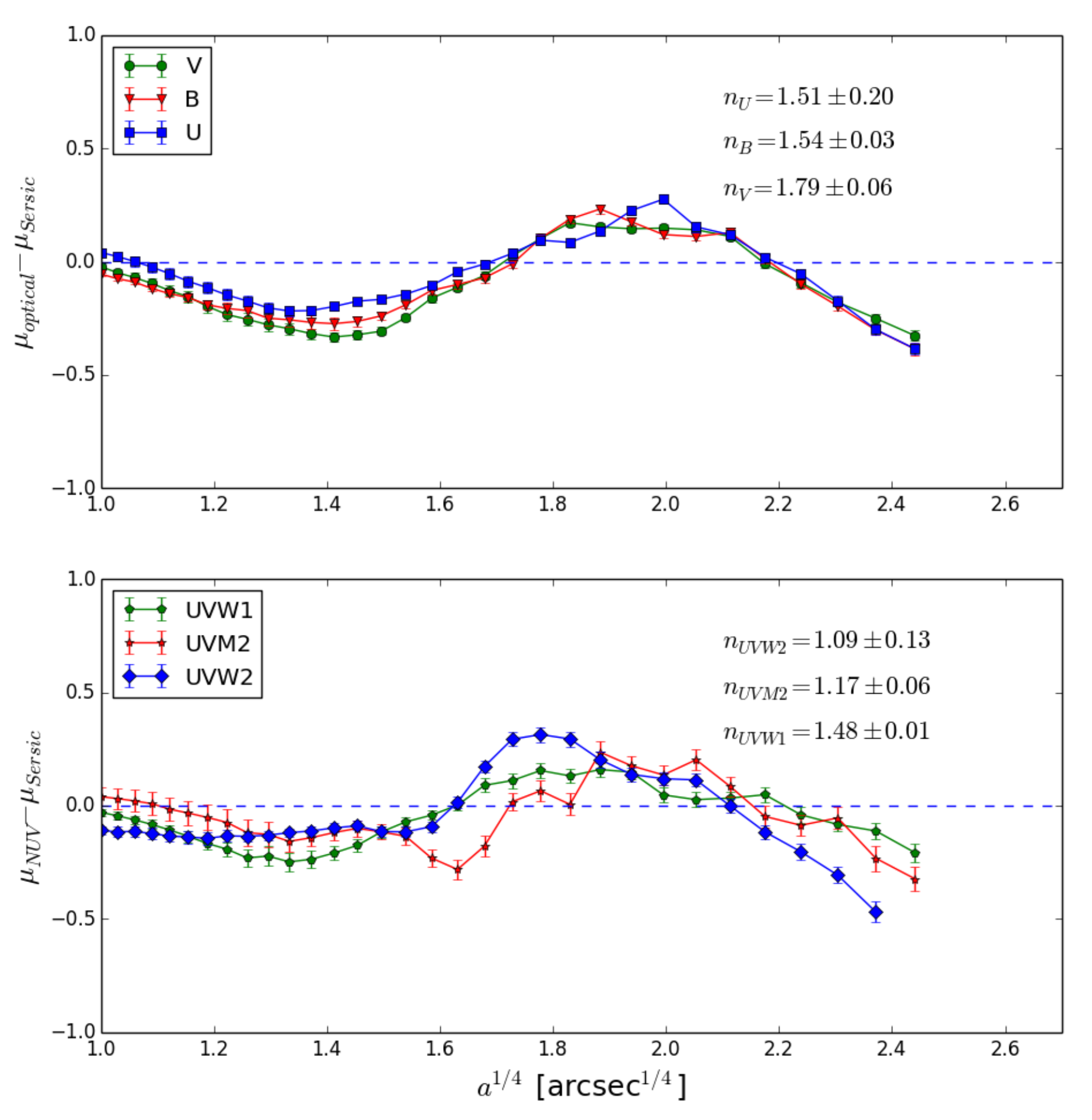}
\caption{Residual from the fit of a single S\'ersic $r^{1/n}$ law of the optical (top panel) and
NUV (bottom panel) luminosity profiles. The value of the S\'ersic indices, for each {\tt UVOT}
band are reported on the top right side of the figure. The values of the indices suggest the presence 
of a disk structure.}
\label{figure3}
\end{figure}
%%%%%%%%%%%%%%%%%%% END FIGURE 3 %%%%%%%%%%%%%%%%%%%%%%%

%--------------------------------------- Table 3 ----------------------------------------------
\begin{table*}
 \centering
   \caption{{\it Swift}-{\tt UVOT} integrated magnitudes} % requires the 
   %topcapt package
   \begin{tabular}{@{} lllllll @{}} % Column formatting, @{} suppresses leading/trailing space
     
      \multicolumn{7}{c}{} \\
%      \cmidrule(r){1-2} % Partial rule. (r) trims the line a little bit on the right; (l) & (lr) also possible
      \hline
      Filter                &   $UVW2$ & $UVM2$ & $UVW1$  & U  &   B  & V \\
      Central $\lambda$    &  2030[\AA] & 2231[\AA] & 2634[\AA] & 3501[\AA] & 4329[\AA]   &  5402 [\AA]    \\
      PSF (FWHM)            & 2\arcsec.92  &  2\arcsec.45  &  2\arcsec.37 & 2\arcsec.37 & 2\arcsec.19  &  2\arcsec.18    \\
      Zero Point $^a$  & 19.11$\pm$0.03 & 18.54$\pm$0.03 & 18.95$\pm$0.03 & 19.36$\pm$0.02 & 18.98$\pm$0.02 & 17.88$\pm$0.01 \\
    Total exp. time   &     1325 [s] & 2255 [s] & 3040 [s] & 652 [s] & 453 [s] & 762[s] \\
    \hline    \hline
     Integrated magnitudes   & [AB mag]   & [AB mag]  & [AB mag] & [AB mag]& [AB mag] & [AB 
       mag] \\
   NGC 454 E    &     17.96$\pm$0.13 & 17.83$\pm$0.25 & 16.61$\pm$0.28 &  15.15$\pm$0.14 & 13.77$\pm$0.18 & 13.20$\pm$0.18    \\
   NGC 454 W   &      15.32$\pm$0.15 & 15.16$\pm$0.13  & 14.90$\pm$0.09 &  
   14.46$\pm$0.12 & 13.51$\pm$0.20 & 13.13$\pm$0.28    \\
   
 \hline
 \label{table3}
   \end{tabular}
 %%%%%%%%%%%%%%%%%%%%%%
%The second row provides the total exposure time, sum of single exposures in each band, in the 
%00035244003 ASDC-ASI product used for our analysis.  
\label{table3} 

\footnotesize{$^a$ provided by \citet{Breeveld11} for converting UVOT count rates to AB mag. \citep{Oke74}.}
\end{table*}
%--------------------------------- end Table 3 ---------------------------------------------------

%--------------------------------------- Table 3Bis ----------------------------------------------
\begin{table*}
\end{table*}
%--------------------------------- end Table 3Bis ---------------------------------------------------

%%%%%%%%%%%%%   FIGURE 4 %%%%%%%%%%%%%%%%%%%
\begin{figure*}
\includegraphics[width=17.5cm]{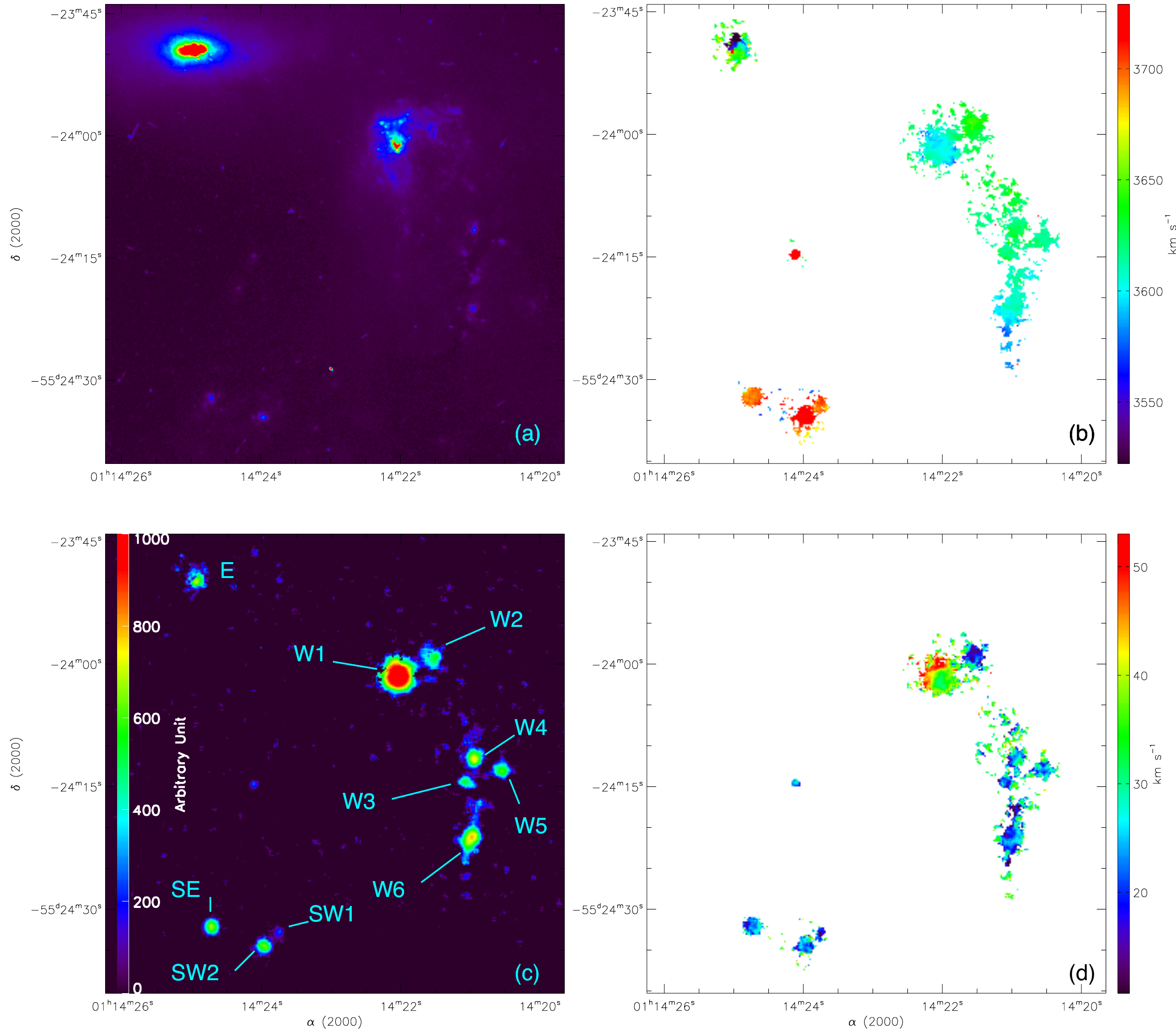}
\caption{(a) HST F450W image of the NGC 454 system \citep{Stiavelli1998}, (b)  2D velocity field of H$\alpha$ emission, 
centred on the NGC 454 systemic velocity, (c) monochromatic H$\alpha$ map,  and (d) H$\alpha$ velocity dispersion map, corrected from broadening.}
\label{figure4}
\end{figure*}
%%%%%%%%%%%%% END FIGURE 4 %%%%%%%%%%%%%%%%%%

%%%%%%%%%%%%%%%%%%%%  FIGURE 5 ELL - W1 & W2 PROFILES REGIONS %%%%%%%%%%%%%%%%%%%%%%%%   
\begin{figure*}
\includegraphics[width=16cm, scale=1.5]{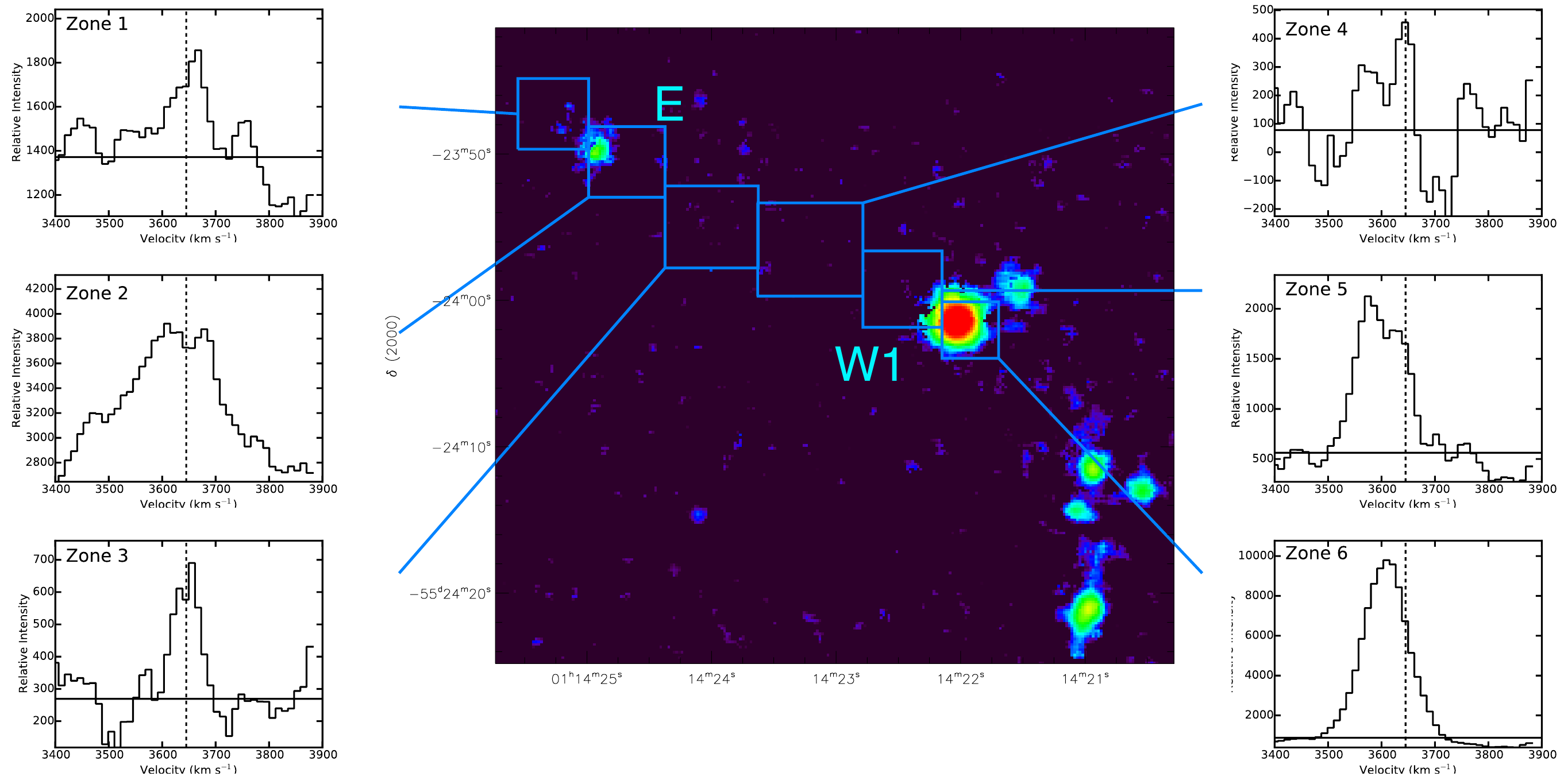}
\caption{Emission profiles in the region NGC~454~E and NGC~454~W1. The vertical
 dotted line indicates the systemic velocity adopted V$_{hel}$=3645 km~s$^{-1}$.The horizontal line indicates the mean continuum level. The map corresponds to the monochromatic emission.}
\label{figure5}
\end{figure*}
%%%%%%%%%%%%%%%%%%%%%%% END FIGURE 5 %%%%%%%%%%%%%%%%%%

%%%%%%%%%%%%%%%%%%%% FIGURE 6 SESW AND W3456 PROFILES REGIONS %%%%%%%%%%%%%%%%%%%%%%%%%%%%%%%
\begin{figure*}
\includegraphics{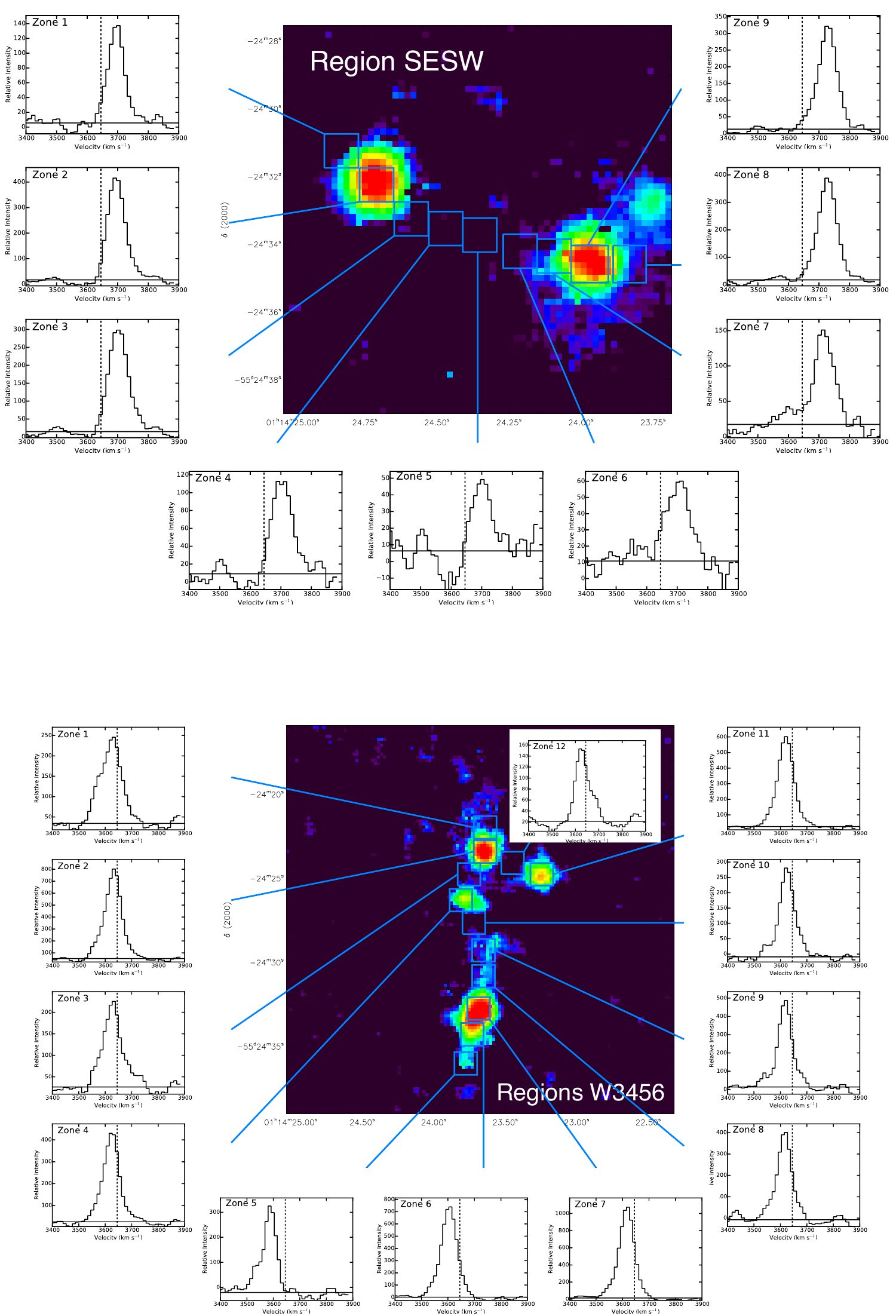}
\caption{Emission profiles  in NGC~454~SE and NGC~454~SW (top panel) and in the 
NGC~454~W region between W3 and W6 complexes (bottom panel). The vertical dotted line indicates
the systemic velocity adopted V$_{hel}$=3645 km~s$^{-1}$.The horizontal line indicates
the mean continuum level. The map corresponds to the monochromatic emission.}
\label{figure6}
\end{figure*}
%%%%%%%%%%%%%%%%%%%%% END FIGURE 6 %%%%%%%%%%%%%%%%%%%%%%%%%%%%%%%

%%%%%%%%%%%%%%%%%%%%% FIGURE 7 - DIAGNOSTIC DIAGRAMS %%%%%%%%%%%%%%%%%%%%%%%%%%%%%%%
\begin{figure*}
\includegraphics[width=16.7cm]{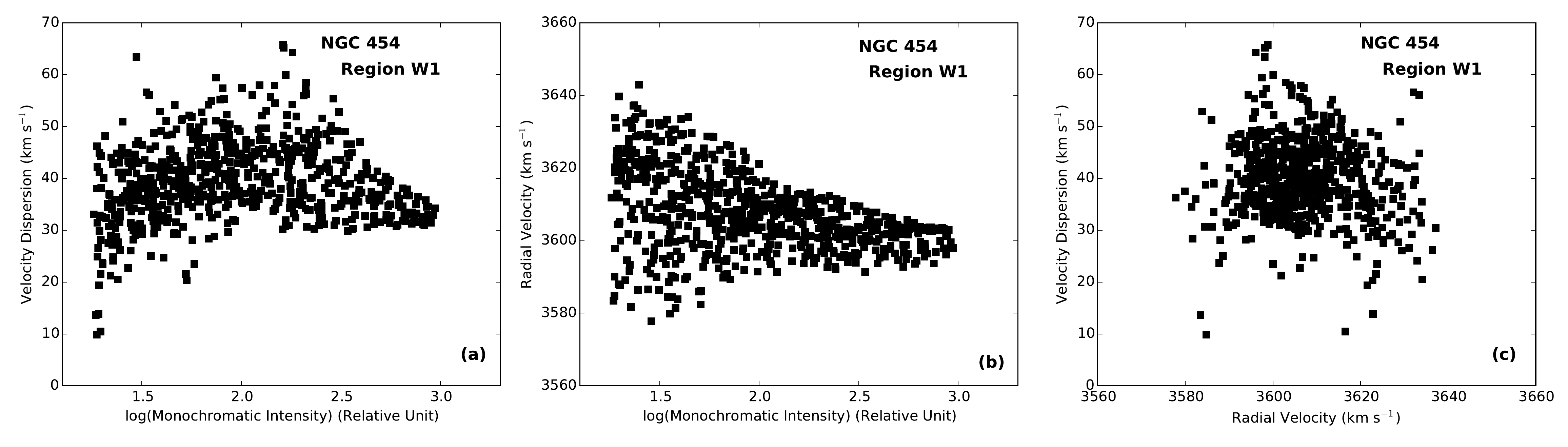}
\includegraphics[width=16.7cm]{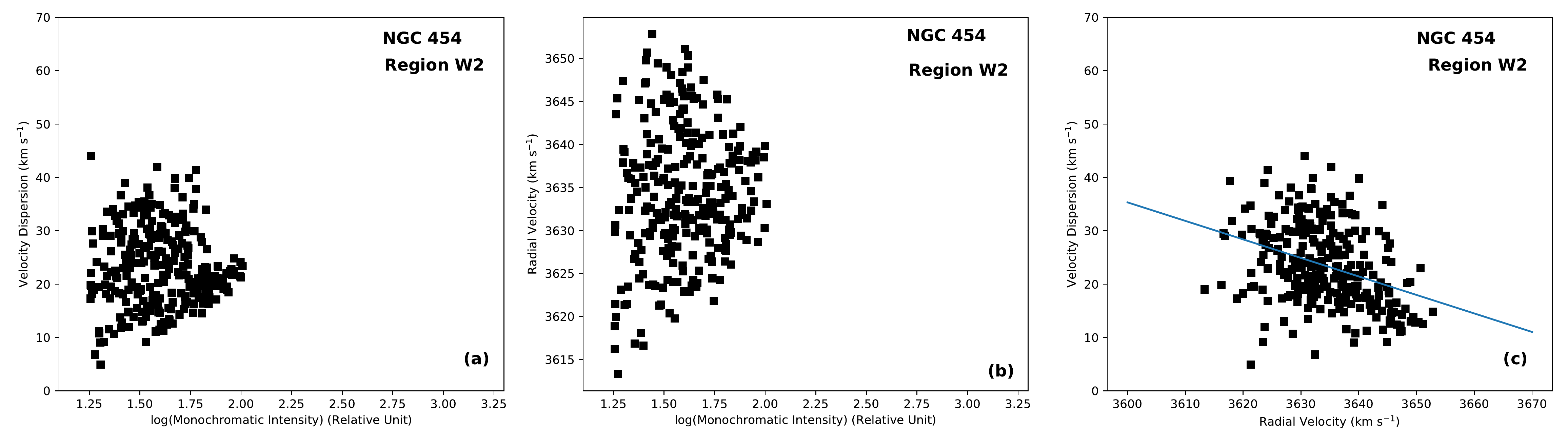}
\includegraphics[width=16.7cm]{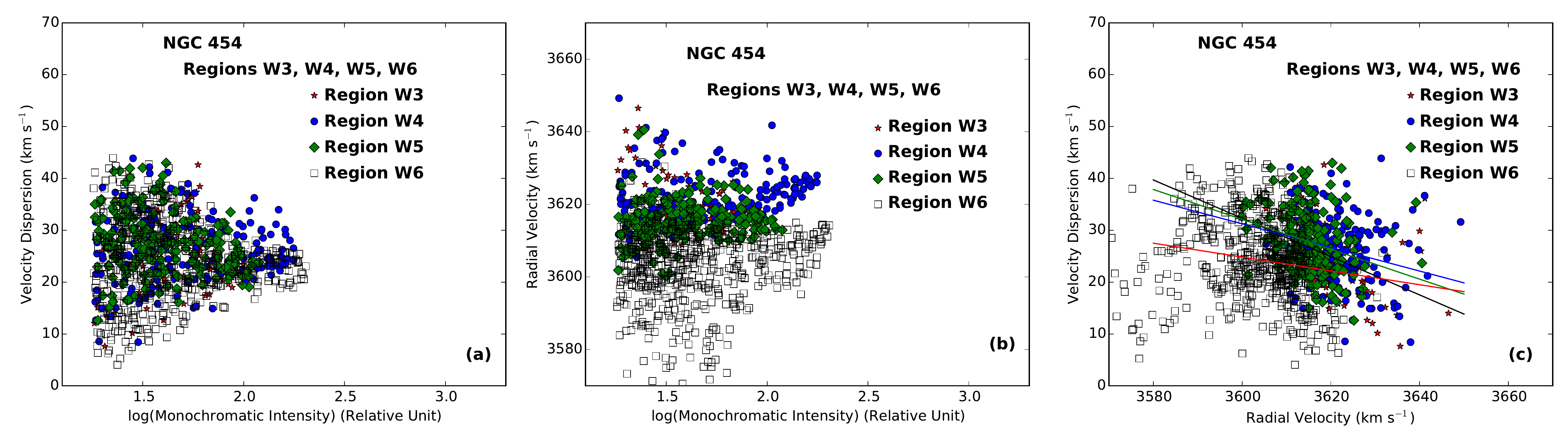}
\includegraphics[width=16.7cm]{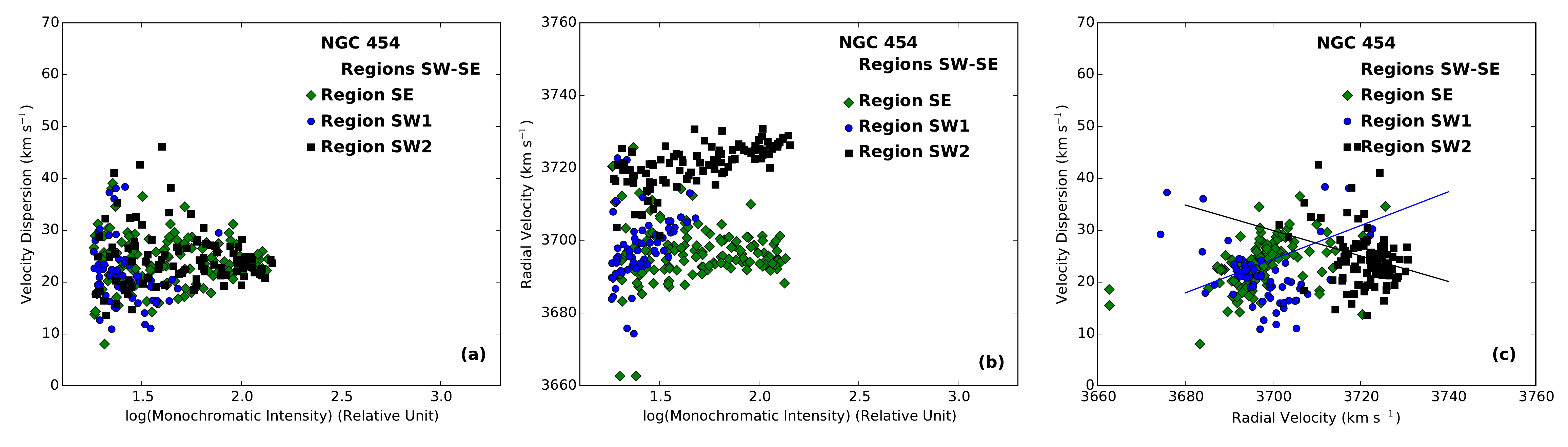}
\caption{From top to bottom: panels plot (a) ($I-\sigma$), panels (b) ($I-V_{r}$),
panels (c) ($V_{hel}-\sigma$) diagnostic diagrams in W1-W6 complexes in NGC~454~W 
and in the NGC~454~SE and SW complexes.  The plot results from a single Gaussian fit
to the line profile. Solid lines in $V_{r}-\sigma$ diagrams represent the linear regressions 
applied when the Pearson's correlation test is robust.}
\label{figure7}
\end{figure*}
%%%%%%%%%%%%%%%%%%%% END FIGURE 7 %%%%%%%%%%%%%%%%%%%%%%%%%%%%%%%

\section{Surface photometry from {\it Swift}-{\tt UVOT} observations} 
\label{sec:UVOTresults}

The color composite images in optical and NUV bands from the {\it Swift}-{\tt UVOT} 
 observations are shown in the top panels of Figure~\ref{figure2}. 
 Images show that the NGC~454~W emission dominates in the NUV bands.
 In NUV NGC~454~W1- W6 complexes
 appear included in a unique envelope which elongates up to NGC~454~SW and
 SE. In NUV two complexes, already revealed in H$\alpha$ by \citet{Johansson1988},
 are projected between NGC~454~SE and NGC~454~W. NGC~454~S, shown
 in Figure~\ref{figure1}, appears connected to NGC~454~SW.\\
\indent
We derived the luminosity profiles and ($UVM2-V$) and ($B-V$) color profiles of NGC~454~E. 
They are shown in the middle and bottom panels
of Figure~\ref{figure2}. Luminosity profiles have been derived using 
the task {\tt ELLIPSE} in the package {\tt IRAF} \citep{Jedr87}. 
These are not corrected for galactic extinction. 
Since we aim at parameterizing the galaxy structure we have 
truncated the profile at $\approx$35\arcsec\  (a$^{1/4}$ = 2.43) where the distortion by NGC~454~W,
in particular in NUV, becomes dominant.\\
\indent
The ($B-V$) color profile tends to become bluer with
the galacto-centric distance as shown in \citet{Stiavelli1998}. The trend
is much clear along the ($UVM2-V$) color profile. 

\citet{Stiavelli1998} parametrized the HST luminosty profiles with
a composite bulge plus disk  model. Due to our poorer resolution and PSF,
to parametrize the trend of optical and NUV surface brightness profiles  we adopt
a S\'ersic r$^{1/n}$ law \citep{Sersic68}, widely used for early-type galaxies 
as a generalization of the r$^{1/4}$ \citet{deVauc48} law \citep[see e.g.][and references 
therein]{Rampazzo2017}. 
We best fit a S\'ersic law convolved with a PSF, using a custom {\tt IDL} routine
based on the  {\tt MPFit} package \citep{Markwardt2009},
accounting for errors in the surface photometry.   
The PSF model is a Gaussian of given FWHM and the convolution is computed using 
FFT on oversampled vectors. We use the nominal value of the FWHM of the PSF of 
the {\tt UVOT} filters.  The residuals, $\mu - \mu_{Sersic}$, are shown in two panels of 
Figures~\ref{figure3}, together with the values of the S\'ersic indices for each of the
{\tt UVOT} bands reported in the top right corner of the two panels. The S\'ersic
indices are in the range 1.09$\pm$0.13$\leq n \leq$1.79$\pm$0.06

We remind that the S\'ersic law has three special cases when $n$=1, the value for an 
exponential profile, and $n=0.5$, for a Gaussian luminosity profile and n=4 for a bulge. The range of our 
S\'ersic indices suggests that NGC~454~E has a disk. Residuals in Figure~\ref{figure3}
shows a trend starting at about $a^{1/4}\simeq1.6 \arcsec$ consistently with a clear change in color
in the ($UVM2-V$) color profile. 

\section{Ionized gas moment maps}
\label{sec:FPresults}
 
We extract from SAM+FP observations the monochromatic 
H$\alpha$ emission map, the radial velocity
and velocity dispersion maps. In Figure~\ref{figure4}  we show
HST F450W image  \citep{Stiavelli1998} (top left panel) on the same
scale with our H$\alpha$ monochromatic map (bottom left panel),   
heliocentric radial velocity map (top right panel),
and  velocity dispersion map (bottom right panel) of the system, corrected from broadening. 

\subsection{H$\alpha$ monochromatic intensity  map}

Both narrow band imaging \citep{Johansson1988} and spectroscopy
\citep{Donzelli2000,Tanvuia2003}  revealed 
H$\alpha$ emission in the NGC 454 system.  
The bottom left panel of Figure~\ref{figure4}
shows the H$\alpha$ monochromatic intensity ($I$) map detected by
 our FP observations.  H$\alpha$ emission is revealed
 in the nucleus of  NGC~454~E,  in the NGC 454~W region
and in the NGC~454~SE and SW complexes. 
NGC~454~W region shows a structured H$\alpha$ emission. In this
region we spot 6 complexes we labeled from W1 to W6, roughly from North
to South, as shown in Figure~\ref{figure4} (bottom left panel).

NGC454~E has a Seyfert~2-type nucleus with broad, 250 and 300 km~s$^{-1}$,
emission line profiles \citep{Johansson1988}. 
That fits into the Free Spectral Range of our etalon (almost 500 
km~s$^{-1}$). We apply a 5$\times$5 px (0.212$\times$0.212 kpc) boxcar 
smoothing to enhance the signal in the nuclear zone. The emission
extends 3\farcs3 $\times$ 2\farcs0 (0.8 kpc $\times$ 0.7 kpc) around the center of NGC~454~E.
Integrating the signal within boxes (see Figure~\ref{figure5})  along the line connecting the NGC~454~E nucleus to the brightest 
complex of NGC~454~W (we labeled W1) we reveal emission lines 
whose profiles have complex shapes. 

The W1 complex  is the brightest in NGC~454~W, showing 
a nearly circular shape  and a diameter of  5\farcs7 (1.3 kpc);  
the W6 complex is the larger, extending  to 8\farcs5 (2 kpc).
%There is a correspondence between the H$\alpha$ map  and our NUV observations \citep[see also][]{Johansson1988}. 
 In the NUV map we may distinguish all the W1--W6 complexes 
but they appear more extended and interconnected than shown
by our FP observations. 
%due to our high resolution and the relatively high background noise. 

The NGC~454~SW and NGC~454~SE emission complexes shown
in Figure~\ref{figure4} are relatively weaker than the W1--W6
complexes. We need to integrate the signal within 5$\times$5 pixels 
bins to detect a connection between these complexes as shown in the 
Johansson's H$\alpha$ narrow band image. 
The bottom panels of Figure~\ref{figure6}  show emission lines profiles in several areas of NGC~454~W, NGC~454~SE and NGC~454~SW.

Both the H$\alpha$ image by \citet{Johansson1988} and in our NUV images
show an elongated emission region between NGC~454~SW and NGC~454~W1
complex, weaker than the other regions. We also detect emission in
between the SW,  SE complex and the SW1 region (see
Figure~\ref{figure4}). Our observation is showing a much smaller
extension (1\farcs1 corresponding to 0.3 kpc) compared to
\citet{Johansson1988}.\\ 
\indent 
To summarize, the W1--W6 as well as SW
and SE in  NGC~454~W are huge (up to 2 kpc wide) complexes of ionized
interstellar medium (ISM hereafter).  Ionized ISM is also found in
NGC~454~E center and along the line connecting it to NGC~454~W1 complex.

\subsection{Radial velocity map}
Figure~\ref{figure4} (top right panel) shows the 2D radial velocity, $V_r$, map of NGC~454. 
 NGC~454~E velocity  field is difficult to interpret because this object is an AGN
showing large line  profiles, almost covering our free spectral range. Nevertheless, 
a velocity gradient  of 130 km~s$^{-1}$, across 4\arcsec (0.94 kpc), is measured. 

Velocities in the NGC~454~W range over 70 km~s$^{-1}$, from maximum of 3645 km~s$^{-1}$ measured
in the W2 complex to a minimum of 3575  km~s$^{-1}$ in the southern tip of W6 complex. 
None of the W1--W6 complexes has a rotation pattern. 
The W6 complex shows a velocity gradient of 35
km~s$^{-1}$ across a length of 8\farcs3  (1.95 kpc). The NGC~454~SW and 
NGC~454~SE complexes do not present velocity gradients. 
With respect to the W1-W6 complexes the SW and SE complexes
are receding with a systemic velocity of 3725 km~s$^{-1}$ and 
3691 km~s$^{-1}$, respectively, with a  velocity difference 
$\Delta V \simeq$115-145 km~s$^{-1}$ with respect to W1 complex.

\subsection{Velocity dispersion map}
The velocity dispersion, $\sigma$,  map is shown in the bottom right panel of Figure~\ref{figure4}.  As mentioned before, NGC~454~E shows very broad profiles, very close to our free spectral range. In those conditions, we prefer not to show a velocity dispersion map for this galaxy.
We measure $\sigma$ values ranging from 14 to 42 km~s$^{-1}$  in the W3, W4, W5, W6 complexes. $\sigma $ values of NGC~454~SW and NGC~454~SE are homogeneous, between 20 and 29 km~s$^{-1}$ .  We point out that values exceeding 10 - 20 km~s$^{-1}$ indicate supersonic motions \citep[see e.g.][]{Smith1970}. The W1 complex shows the highest values, up to 66 km~s$^{-1}$, while in the W2 complex, $\sigma$  ranges from 13 to 32 km~s$^{-1}$. 
%The comparison
%between  the H$\alpha$ mono-chromatic image and the velocity dispersion 
%map  (Figure\ref{figure4}, bottom panels, left and right respectively)
%evidences that as $\sigma$ is higher as the  intensity ($I$) is lower
%and viceversa as typical in H\,{\textsc{ii}}  regions \citep[see
%e.g.][]{Munoz-Tunon1996}.\\

\subsection{Emission between H\,{\small \text{II}} regions}
\label{subsec:Emission between}

Figure~\ref{figure5} and Figures~\ref{figure6} show several regions in between NGC~454~E and NGC~454~W1 (Figure~\ref{figure5}), NGC~454~SW and NGC~454~SE (Figure~\ref{figure6} top) and between NGC~454~W3~W4~W5 and W6 (Figure~\ref{figure6} bottom). 

Figure~\ref{figure5}, regions $1$ and $2$ show the emission centered in the elliptical. As mentioned before, the center of NGC~454~E has a broad emission as shown in region $2$. With an emission peak above 3800 (in relative units) and a background of 2500, region $2$ has a very high intensity considering that background is dominated by poisson noise and even if we consider that the profile has a FWHM of 300km~s$^{-1}$ (compared to the 500 km~s$^{-1}$ of the free spectral range), we still see the emission of the center of the elliptical. 
Regions $3$ and $4$ show areas in between NGC~454~E and NGC~454~W1, when region $3$ shows a confortable emission line,   region $4$ shows the limit of our detection with an emission peak at one  $\sigma$ above the continuum level. Regions $5$ and $6$ show NGC~454~W1 emission, a detailed discussion about this latest is given in  subsection ~\ref{sec:W1statistics} and section~\ref{sec:Profile decomposition}. \\

Figure~\ref{figure6} (top) shows emission profiles between the SE and SW regions. Emission is very clear across both regions. No substantial velocity gradient is visible even if the radial velocity of SW is 30 km~s$^{-1}$  lower than SE. \\

Figure~\ref{figure6} (bottom) shows the connection between the remaining regions (NGC~454~W3 to W6). Emission is strong and it is clear that all regions are connected. Zones $1$ and $12$ show asymmetric profiles, probably due to
a second component. A radial velocity gradient is visible between the southern tip with zone $5$ and region $1$.

 \begin{figure}
 \includegraphics[width=8.5cm]{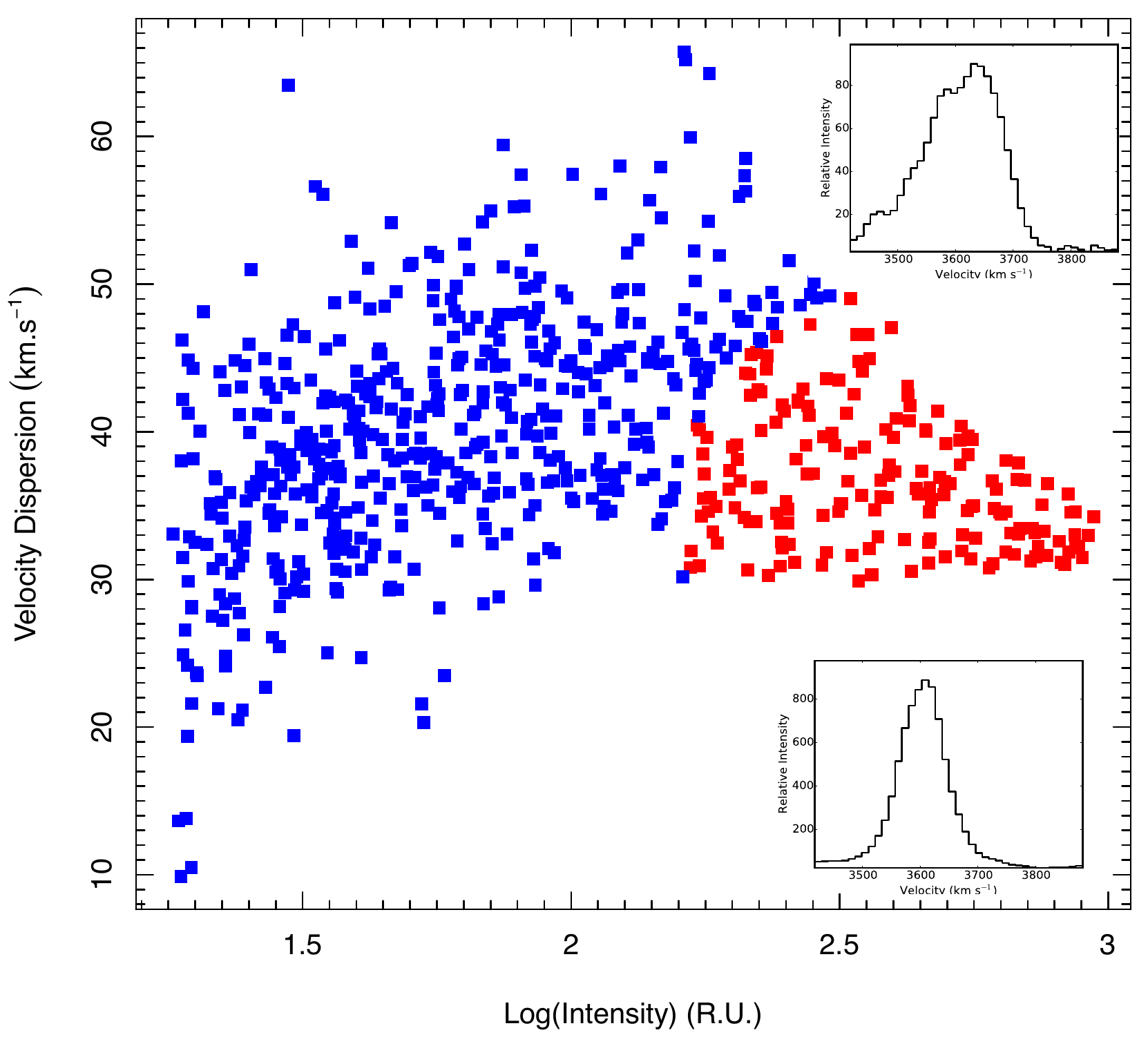}
 \includegraphics[width=8.5cm]{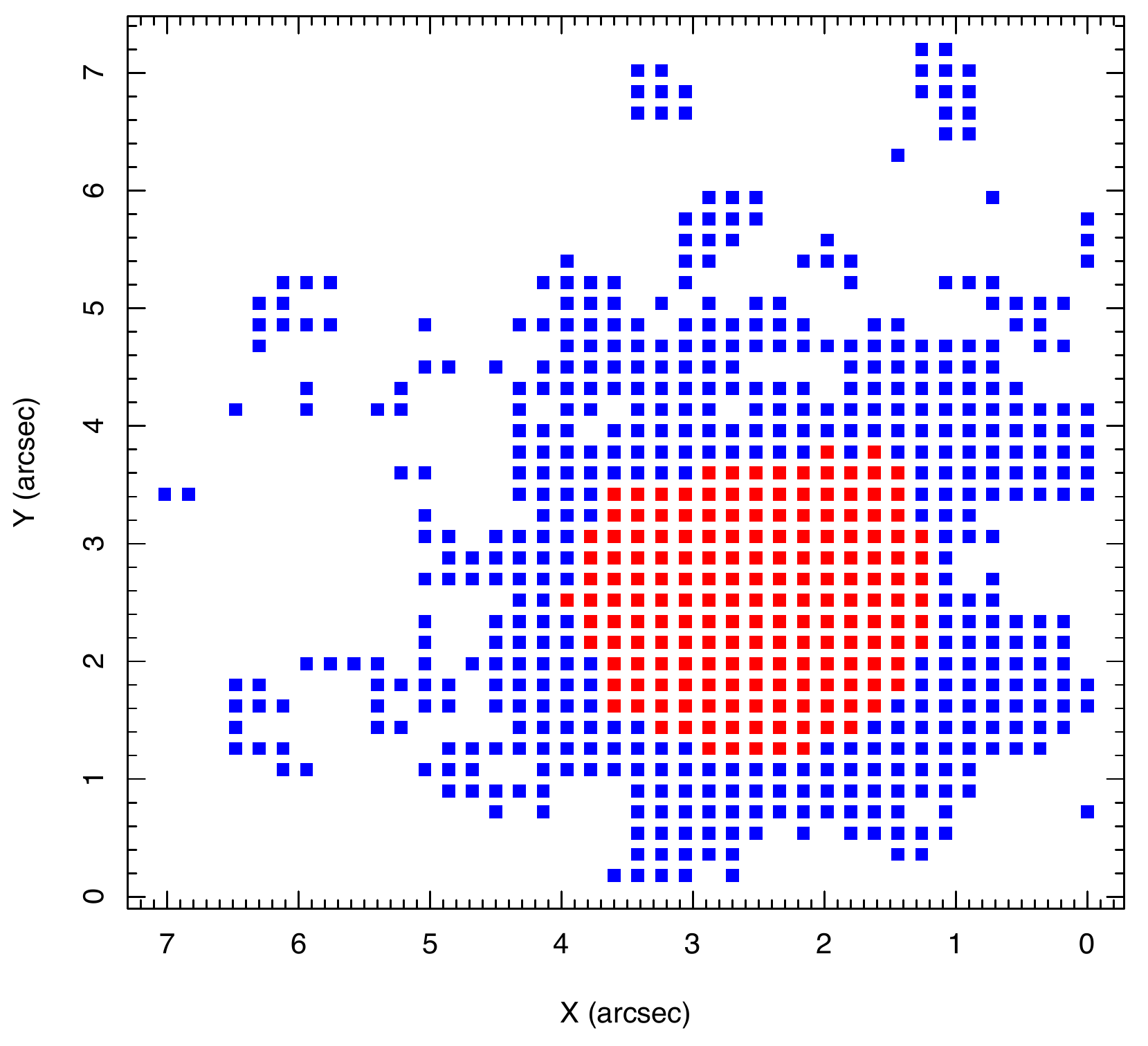}
 \caption{(bottom panel) The 3$\times$3 pixels sampling of the NGC~454~W1 complex.
 (top panel) ($I-\sigma$) plot. The colors highlight two regions, the central and 
 outskirts, with different regimes (see Section~\ref{sec:W1statistics} for details).  Top profile shows a typical monochromatic emission representative of the blue points, and bottom profile representative of the red points.}
 \label{figure8}
 \end{figure}
%%%%%%%%%%%%%%%%%%%%%% END FIGURE 8%%%%%%%%%%%%%%%%%%%%%%%%%%%%%%%%%%%%%%%%%%%%%%%%%

\section{H\,{\small \text{II}} regions diagnostic diagrams}
\label{sec:Diagnostic diagrams}

\subsection{Description of the complexes}
\label{subsec:Descriptions}

Complexes in NGC~454~W, share  similar dynamical characteristics
both with  Giant H\,{\textsc{ii}} Regions and the so-called
H\,{\textsc{ii}} Galaxies. Firstly, like GH\,{\textsc{ii}}Rs, W1
has high supersonic profiles \citep{Smith1970} and, secondly, the high
velocity dispersion surrenders high monochromatic emission. Several
studies using Fabry-Perot interferometer \citep{Munoz-Tunon1996} found
this signature in nearby Giant H\,{\textsc{ii}} Region, like NGC
604. More recently, still using Fabry-Perot and IFU spectroscopy
\citet{Bordalo2009, Moiseev2012, Plana2017}, 
evidences have been found of such signature within dwarf H\,{\textsc{ii}}  galaxies.\\
\indent
Figure~\ref{figure7} considers three different diagnostic diagrams used
to study the kinematics of  H\,{\textsc{ii}} regions 
%\citep[see e.g.][and references therein]{Munoz-Tunon1996,Bordalo2009, Moiseev2012}.
These diagnostic diagrams are shown for NGC~454~W complexes (W1-W6
identified in Figure~\ref{figure4}) and for NGC~454~SW and SE regions. \\
\indent
%These plots show that the W1 region has a different behaviour compared to other W complexes.
The W1 complex has the larger intensity range.  We will discuss the ($I-\sigma$) regimes in this region in Section~\ref{sec:W1statistics} with a statistical approach.\\

The panels (a) in Figure~\ref{figure7} represent the intensity vs the dispersion velocity (sigma). The W2 -- W6 as well as SW1, SW2 and SE complexes show similar intensity and sigma ranges. %The W2 complex
In W3 -- W6 as well as NGC~454~SW and SE complexes the scatter of $\sigma$ increases as the intensity decreases. As discussed by \citet{Moiseev2012} (their Figure 6) this  shape is produced within star forming complexes with significant excursion of gas densities, indicating low density, turbulent ISM. At  high density (high intensity) regime H\,{\textsc{ii}} regions have either nearly constant or low scatter  $\sigma$. However, towards lower density (low intensity) the
high perturbed/turbulent gas surrounding H\,{\textsc{ii}} regions may emerge so decreasing the intensity the $\sigma$ scatter may increase. We have separated the different regions: W3, W4, W5 and W6 as SE, SW1 and SW2  complexes in these diagrams. The last row of Figure~\ref{figure7} shows the SE, SW1 and SW2 complexes I vs $\sigma$ diagram, but none of the three complexes shows a different pattern.  Introduced by \citet{Munoz-Tunon1996}, the ($I-\sigma$) diagram has been used by those authors to identify expanding shells by localising inclined bands.  This interpretation is based on the fact that the velocity dispersion should be higher at the center of the shell and the intensity
lower because less material is crossed along the line of sight than at the shell inner and outer edges. Assuming this pattern, the inclination of the band can also be interpreted in term of age of the shell itself. As the shell ages, the velocity dispersion at the center decreases as well as the intensity difference between the center and inner edge of the shell \citep[see figure 3 in][]{Munoz-Tunon1996}. The \citet{Moiseev2012} interpretation tends to act for larger scales, where high velocity dispersion is not related to specific expanding shells, but rather belong to the diffuse low brightness emission.\\
\indent

The panels (b) in Figure~\ref{figure7} represent the intensity vs.
radial velocity, ($I-V_r$), diagram of  W1-W6 complexes in NGC~454~W,
and in the SW1, SW2 and SE. 

%\citet{Bordalo2009} presented idealized pattern for this diagram as a way to interpret it. 

\indent
The range of radial velocity, $V_r$, within complexes is small, of the order of 20-40
km~s$^{-1}$ in the  W3 -- W6 and NGC~454~SW1, SW2 and SE complexes.
Large radial velocity excursion at all intensities is found in W2.
% while W3 -- W6 complexes, shows almost similar variations, 30-40 km~s$^{-1}$, with the monochromatic intensity. 
The  W3, W4, W5 and W6 complexes show different radial velocities, with W4 having the highest and W6 the
lowest. W6 velocity range is larger because of the velocity gradient we already mentioned in Section~\ref{sec:FPresults}. The separation in
radial velocity between the SE, SW1 and SW2 complexes is shown in the plot (b) of the last row of Figure~\ref{figure7}. The complex
NGC~454~SE, SW and SW2 have a small variation, 10-20 km~s$^{-1}$, of $V_r$
while the monochromatic intensity range is similar to W3 -- W6 complexes. 
It also appears that SW1 complex has a closer radial velocity with SE than with SW2. 
According to \citep{Bordalo2009} (their figure 13b), a vertical band in this diagram, representing a 
velocity variation in a short intensity range, means a radial motion such as an expansion,  but it 
could also means an inflow. The physical
mechanisms in action are several including turbulence, winds, flows, bubbles  or the self gravity of the complexes at different scales. Even if a vertical band can appear in the plot representing W2 (second row), it is difficult to interpret it as a signature of a radial motion, the intensity range being too wide.\\
\indent
The panels (c) in Figure~\ref{figure7} consider the ($V_r-\sigma$)
diagrams in the same regions. \citet{Bordalo2009} pointed out that a
dependence between the variables may indicate systematic relative motion
of the clouds in the complex. They presented an idealized pattern for
this diagram. Inclined  ($V_r-\sigma$) patterns would represent
systematic motion like Champagne flows, such that cloud of gas with high
$\sigma$ moves away from us (positive slope) or toward us (negative
slope).

We perform a standard Pearson's product-moment correlation test for
the different complexes of NGC~454, in order to show the existence of
systematic motions mentioned above. Except for W1 and SW1, all regions
show weak-moderate correlation, according this test. W2 has a correlation
coefficient of -0.34, W3 of -0.43, W4 of -0.23, W5 of -0.27, W6 of
-0.17. The SE  and SW2 regions also have a weak-moderate correlation with a
coefficient of 0.48 and -0.25.  All of them with a 99.9\% confidence
level. The case of SW1 is a bit more complex. The Pearson test is not
conclusive and we decide to use a robust correlation  test in order to
put lower weight in marginal points. Using the {\tt WRS2} package in {\tt R}, we
found a correlation of -0.2 with a 90\% confidence level.\\
\indent
We perform a simple linear regression (the solid line in
Figure~\ref{figure7} panels (c)) for the regions where the Pearson test
shows a weak-moderate correlation: W2 to W6, SE and SW2.  
%According to  \citet{Bordalo2009}, a correlation between $V_r$ and $\sigma$ is a signature of a systematic motion. 
In this context W2 and SW2 regions can be
interpreted as complexes with relatively high dispersion, moving toward
the observer (negative slope). In the case of the SE complex, the slope
is positive and it can be interpreted as a complex moving away from the
observer. 

Previous studies, on Giant H\,{\textsc{ii}} Region or emitting dwarf galaxies, also used those diagnostic diagrams when these objects are smaller and the scale resolution much smaller than here.
For example NGC 604 study from  \citet{Munoz-Tunon1996} or dwarf galaxies from \citet{Moiseev2012} or \citet{Bordalo2009} have respectively scale resolution of 3.31pc/\arcsec and 21pc/\arcsec. This is ten times smaller than
our object. Even though, we found remarkable similarities between diagnostics diagrams.

%%%%%%%%%%%%%%%%%%%%% FIGURE 9 - PROFILES DECOMPOSITION %%%%%%%%%%%%%%%%%%%%%%%%%%%
 \begin{figure*}
 \includegraphics[width=21 cm,scale=1.6, angle=90]{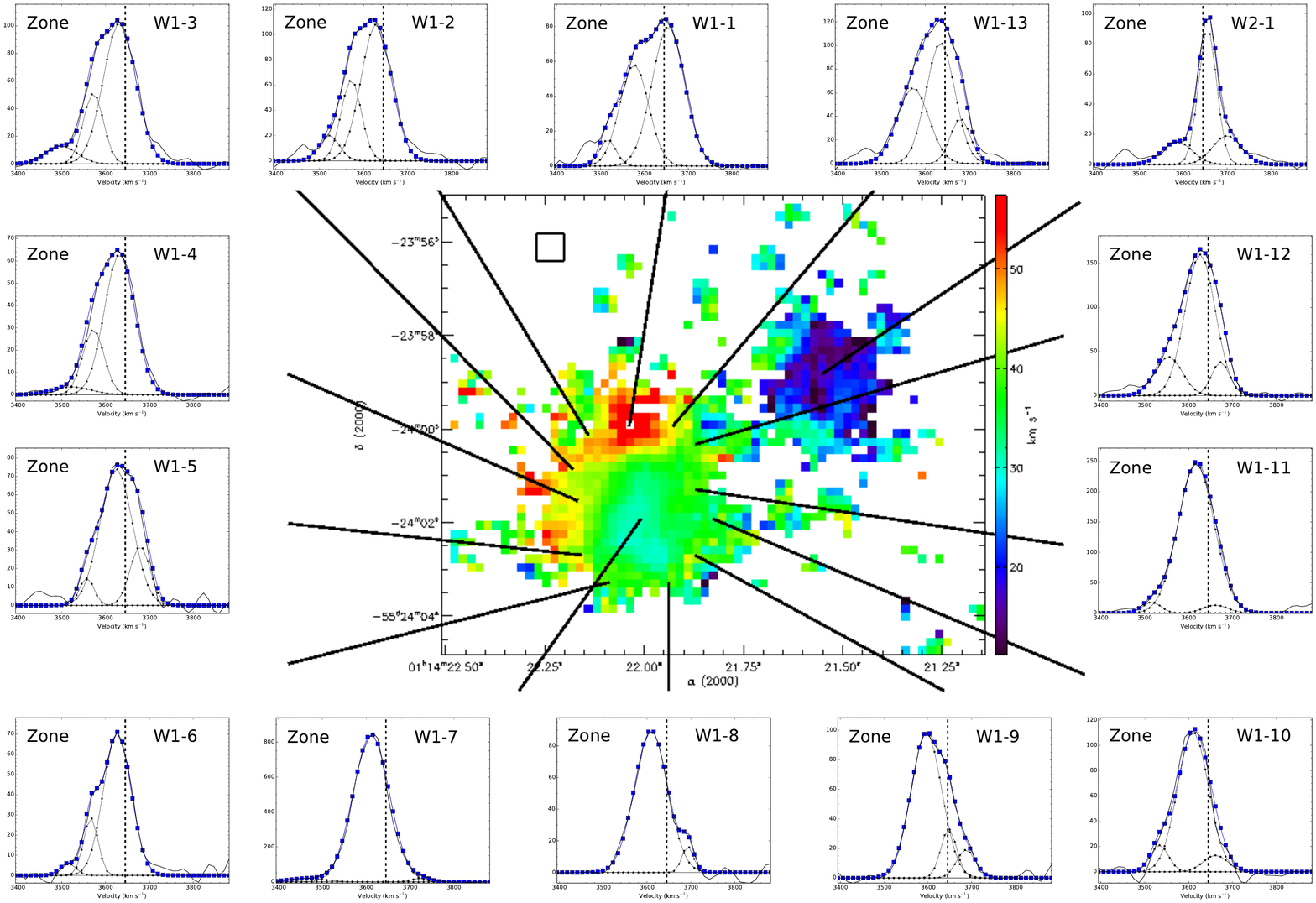}
 \caption{Integrated profiles in NGC 454~W1 and W2. Three Gaussian components fit of the emission lines in 
 NGC~454~W1 and W2 complexes. Results of the fitting are reported in Table~\ref{table3}. The vertical dotted line indicates the systemic velocity adopted V$_{hel}$=3645 km~s$^{-1}$.The horizontal line indicates
 the mean background level. The small square approximatively represents the size of the area of the integrated profiles (3 $\times$ 3px or 0.54\arcsec $\times$ 0.54\arcsec). The map corresponds to the dispersion velocity field, corrected
 from broadening.}
 \label{figure9}
 \end{figure*}
%%%%%%%%%%%%%%%%%%%%%% END FIGURE 9 %%%%%%%%%%%%%%%%%%%%%%%%%%%%%%%%%%%%%%%%%%%%%%%%

%%%%%%%%%%%%%%%%%%%%%% FIGURE 10 - DIAGNOSTIC DIAGRAM W1 + MAIN COMPONENT%%%%%%%%%%%
  \begin{figure}
 \includegraphics[width=8.5cm]{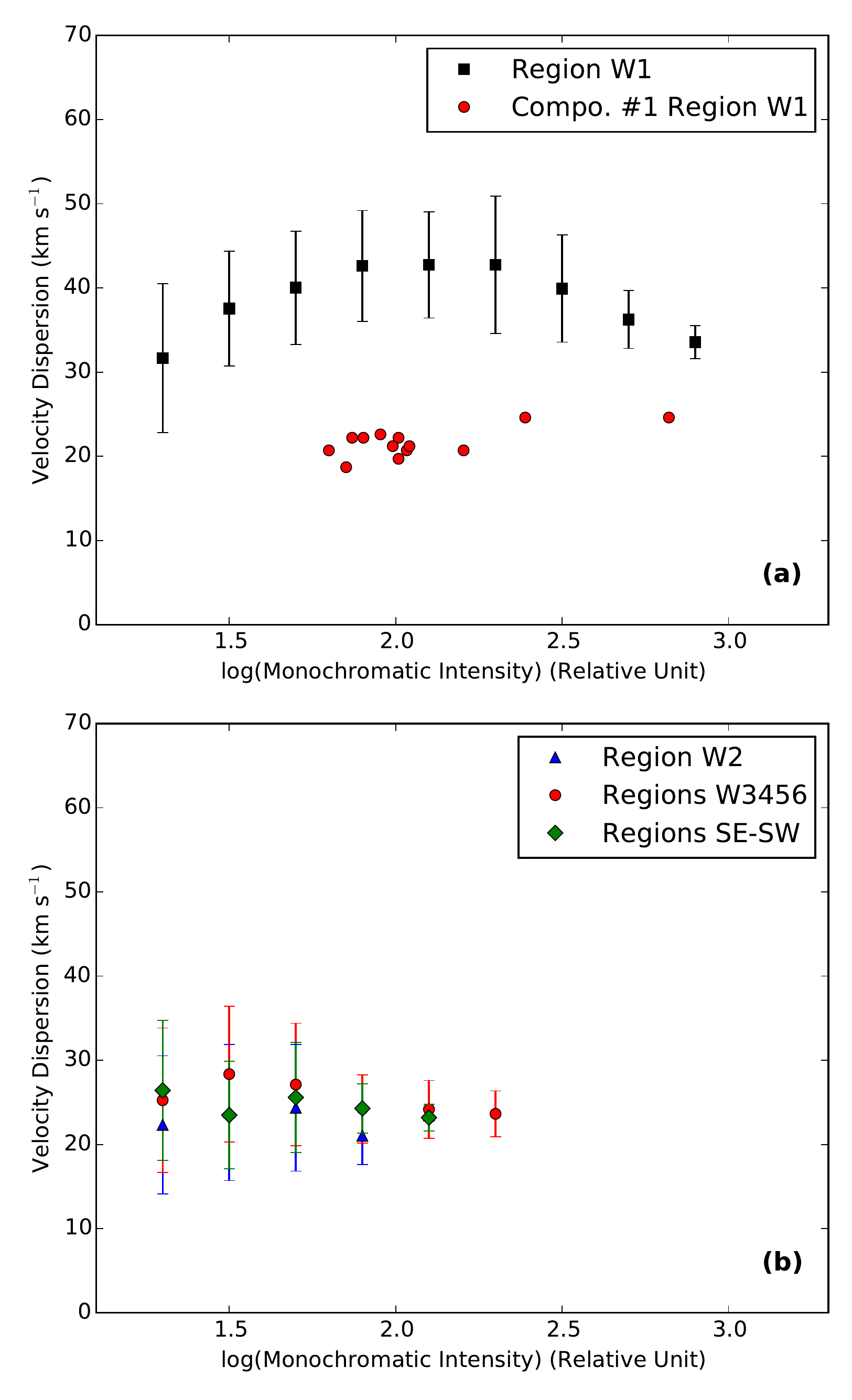}
\caption{($I-\sigma$) diagnostic diagram for the W1 complex (top panel)
and for the other complexes in NGC~454~W (bottom panel).The $\sigma$ values are averaged
over intensity bins. Red dots on W1 complex show the main component in line
profile decomposition shown in Figure~\ref{figure9}.  } 
 \label{figure10}
 \end{figure}
%%%%%%%%%%%%%%%%%%%%%% END FIGURE 10 %%%%%%%%%%%%%%%%%%%%%%%%%%%%%%%%%%%%%%%%%%%%%%%%%

%--------------------------------------- Table 4 ----------------------------------------------
   
   \begin{table*}
   % \begin{minipage}[t]{\columnwidth}
   \caption{H$\alpha$ line profiles Gaussian components in Region W1}
   \label{table4}
   \centering
   \renewcommand{\footnoterule}{} % to avoid a line before footnotes
    \begin{tabular}{l|ccc|ccc|ccc}
    \hline \hline
   Regions & \multicolumn{3}{c}{Component 1} & \multicolumn{3}{c}{Component 2} & 
   \multicolumn{3}{c}{Component 3} \\
   & I & V$_r$ & $\sigma$ & I & V$_r$ & $\sigma$ & I & V$_r$ & $\sigma$ \\
   &R.U. & [km~s$^{-1}$] & [km~s$^{-1}]$ & R.U. & [km~s$^{-1}$] & [km~s$^{-1}$] & R.U. & [km~s$^{-1}$] & [km~s$^{-1}$] \\
   (1) & (2) & (3) & (4) & (5) & (6) & (7) & (8) & (9) & (10) \\
   \hline
    Region W1~1 &  80.0 & 3654 & 22.2 & 58.0 & 3577 & 19.7 & 15.0 & 3516 & 11.3 \\
   
   Region W1~2 & 108.0 & 3629 & 20.7 & 64.0 & 3572 & 13.8 & 20.0 & 3522 & 13.8 \\
   
   Region W1~3 & 102.0 & 3632 & 22.2 & 51.0 & 3572 & 13.8 & 13.0 & 3499 & 21.2 \\
   
   Region W1~4 & 063.0 & 3632 & 20.7 & 29.0 & 3572 &  4.8 & 03.5 & 3522 & 27.1 \\
   
   Region W1~5 &  74.0 & 3624 & 22.2 & 32.0 & 3679 & 13.8 &15.0 & 3557 & 09.9 \\
   
   Region W1~6 &  71.0 & 3627 & 18.7 & 28.5 & 3566 & 9.9 & 6.0 & 3514 & 10.1 \\
   
   Region W1~7 & 662.0 & 3611 & 24.6 & 25.0 & 3487 & 26.6 & 20.0 & 3725 & 22.1 \\
   
   Region W1~8 &  90.0 & 3609 & 22.6 & 16.5 & 3693 & 7.9 &  0.0 & 3383 & 00.0 \\
   
Region W1~9 &  98.0 & 3596 & 21.2 & 33.0 & 3647 & 10.9 & 19.0 & 3687 & 12.9 \\
   
   Region W1~10 & 110.5 & 3610 & 21.2 & 21.0 & 3534 & 12.3 & 13.0 & 3661 & 16.8 \\
   
   Region W1~11 & 245.5 & 3617 & 24.6 & 18.0 & 3516 & 12.3 & 13.0 & 3661 & 16.8 \\
   
   Region W1~12 & 160.0 & 3627 & 20.7 & 44.0 & 3555 & 18.7 & 38.0 & 3673 & 12.4 \\
   
   Region W1~13 & 102.0 & 3635 & 19.7 & 64.0 & 3572 & 21.7 & 38.0 & 3680 & 13.8 \\
   
   Region W2~1 &  90.0 & 3656 & 12.3 & 19.0 & 3690 & 19.7 & 15.0 & 3586 & 20.7 \\
   \hline
   \hline
    \end{tabular}
    
  % \end{minipage} 
  The intensity, I, (col.s 2, 5 and 8) is in relative units (R.U.).
   \end{table*}
   %--------------------------------- end Table 4 ---------------------------------------------------

\subsection{Statistical analysis of the ($I-\sigma$) diagrams for W1 complex}
\label{sec:W1statistics}
We use the {\tt R} statistical package ({\tt R} Development Core Team
2009) to analyze the ($I-\sigma$) diagram of the W1 complex. {\tt R}  is
largely use in different statistical analysis. The {\tt Mclust} routine
only has been recently used in astrophysics by \citep{Einasto2010} to
detect structure in galaxies clusters. We aim at finding how many
independent components are present (task {\tt Mclust}), to locate them
in the diagram and in the $\sigma$ map (so-called geographic location).
{\tt Mclust} is a {\tt R} function for model-based clustering,
classification, and density estimation based on finite Gaussian mixture
modeling. An integrated approach to finite mixture models is provided,
with routines that combine model-based hierarchical clustering and
several tools for model selection \citep[see][]{Fraley2007}.\\
\indent 
 For a bivariate random sample ${\bf x}$ be a realization from a finite mixture of $m>1$ 
distributions, it should follow

\begin{equation}
p({\bf x}|\pi,\{\mu_k,\Sigma_k\})=\sum_k \pi_k\phi({\bf x}|\mu_k,\Sigma_k),
\end{equation}

\noindent where $\phi$ is the multivariate normal density

\begin{equation}
\phi({\bf x}|\mu , \Sigma)=(2\pi)^{-d/2}|\Sigma|^{-1/2}\exp{\{-{1\over 2}({\bf x} - \mu)^\prime 
\Sigma^{-1} ({\bf x} - \mu)\}},
\end{equation}

\noindent ${\bf \pi}=\{\pi_1,...,\pi_m\}$ are the mixing weights or probabilities
(such that $\pi_k > 0$ and $\sum_k^m \pi_k =1$), $(\mu_k,\Sigma_k)$ are the
 mean the covariance matrix of the component $k$,
and $d$ is the dimension of the data. A central question in finite mixture modeling is how many 
components should be included in the mixture. In the multivariate setting, the volume, shape, and 
orientation of the covariances define different models (or parametrization) with their different 
geometric characteristics.
In {\tt Mclust}, the number of mixing components and the best covariance parameterization are 
selected using the Bayesian Information Criterion (BIC).
The task outputs $\mu_k$, $\Sigma_k$ and $\pi_k$ for $k$ running from 1 to $m$. 
{\tt Mclust} also relates each  element in the dataset to a particular component
in the mixture. To gain some flexibility on this classification, we combine the
central result of {\tt Mclust}, the number of components $m$, with the result of another {\tt R}
task: the {\tt mvnormalmixEM} function.

This task belongs to the {\tt mixtools} package, 
which provides a set of  functions for
analyzing a variety of finite mixture  models. The general methodology
used in {\tt mixtools} involves the representation of the mixture problem as a
particular case of maximum likelihood estimation when the
observations can be viewed as incomplete data. The code uses the 
Expectation-Maximization (EM) algorithm that maximizes the conditional
expected log-likelihood at each M-step of the algorithm -- see details
in \citet{Benaglia2009}. The code returns the posterior probabilities
for each observation with respect to the $m$ different components. \\

Since running {\tt Mclust} results  $m=2$ components, we then use the task
{\tt mvnormalmixEM}, looking at two independent classes with 80\% confidence
in the ($I-\sigma$) maps. Once these two classes have been found, 
we have represented them in the ($I-\sigma$) diagram (Figure~\ref{figure8} 
upper panel) and in the $\sigma$ map (Figure~\ref{figure8} lower panel). The figure
clearly shows the two regions, one in the center (low
dispersion and strong emission) and the other surrounding it (high
dispersion and low emission). W1 is an extended ionized ISM complex
so we cannot easily apply the interpretative scheme of H\,{\textsc{ii}}  regions.
We may exclude that W1 may be interpreted as an expanding 
wind-blown bubble would have a different signature in a
($I-\sigma$) map: $\sigma$ values should decrease from the center to the edge
of the shell, according \citet{Lagrois2009}. The two regimes evidenced by
the statistical approach support the picture proposed by \citet{Moiseev2012}
in which the W1 complex can be viewed as composed of a giant dense HII region in the central part and turbulent low-density gas cloud in its outskirt.

\section{H$\alpha$ line profile decomposition}
\label{sec:Profile decomposition}

Figure~\ref{figure4} (bottom right panel)  shows the large range of $\sigma$ in the W1 complex with respect to the other complexes in NGC~454~W as well as in the NGC~454~SW and SE. Figure~\ref{figure5} shows that emission
lines are quite broad so that the  high velocity dispersion in the W1 complex can be attributed to the presence of multiple components in the emission profiles. Figure~\ref{figure9}
shows line profiles resulting from the mapping of the W1 and W2 complexes.\\
\indent
We perform a gaussian decomposition of the 14 H$\alpha$ line profiles  (13 in W1 and one in W2 as a sort of control field) shown in Figure~\ref{figure9}.  Each region represents a 3$\times$3 pixels box (0.54\arcsec$\times$0.54\arcsec\ area, 0.13$\times$0.13 kpc). The decomposition has been performed using a home made program with three different Gaussian components. We first fit the brightest component, subtracts it and then fit the two others, until the final fit converges. Figure~\ref{figure9} shows the decomposition of these profiles and the location of areas in W1 and W2. Table~\ref{table4} lists the characteristics of the three Gaussian components ordered by decreasing intensity component. \\ 
\indent

%On the other side, a Gaussian decomposition is  uncertain
%since the solution is not unique for a fixed number of components and
%conversely is possible to increase the number of components to improve the fit. 
We are aware that a simple mathematical approach is always unsatisfactory, since the composition is not unique, but linked to a physical and kinematical interpretation, we are reasonably satisfied with the result.\\
\indent
Considering the results of the decomposition reported in Table~\ref{table4},  shown in Figure~\ref{figure9} we draw the following conclusions:

The central region, labeled W1~7, has a symmetrical profile  when compared to all the others regions. The $\sigma$ of the brightest Gaussian component of W1 complex has supersonic values between 20 and 25 $km~s^{-1}$. The $\sigma$ of the main component of all regions in W1 is  larger than in W2.

With respect to the systemic velocity the main Gaussian component in W1~[1 to 5] is red-shifted while in the zones W1~[9 to 11] is blue-shifted, sketching a sort of rotation pattern being at the opposite sides of the W1 complex center. 
We also can note that positions of the second component with respect to the main one (second red-shifted component at one side of the W1 complex, and the second blue-shifted component at the other side)  could be seen as a bipolar outflow due to massive star formation.

Several zones of W1 clearly show profiles with an apparent second component (e.g. W1~1, W1~2, W1~3, W1~4 and W1~6), while, in general, the other zones,  including the W2 zone, need fainter components to fit the wings of the
line profiles.

We conclude that even with multiple Gaussian fit analysis no un-ambiguous rotation pattern emerges in the W1 complex.\\ 
\indent
Using the characteristics of the Gaussian decomposition resumed in Table~\ref{table4}, we present, in Figure~\ref{figure10}, a revised
($I-\sigma$) diagnostic diagram, showing the ($I-\sigma$) diagram for the main i.e. the brightest component. In Figure~\ref{figure10}, top panel, we present the mean $\sigma$ per intensity bin. The associated
error bar represents the standard deviation in the bin. The red points represent the intensity and $\sigma$ of the main component (see component 1 in Table~\ref{table4}). As mentioned before, $\sigma$  is
significantly lower, still largely supersonic and similar, on the average,  to regions W2--W6, and SE-SW shown in the bottom panel of Figure~\ref{figure10}. However, the $\sigma$ of the main component in W1 shows a positive slope with $I$.

\section{Discussion}\label{sec:discussion}

\subsection{NGC 454: General view}
\label{subsec:scenario}

%The NGC~454 system belongs to the so-called mixed morphology pairs of galaxies, i.e. composed of an early and a late type galaxy. Although the vast majority of pair members have similar morphological types, a first
%light on the existence of mixed morphology pairs has been shed by the \citet{Karachentsev1972} Catalog of Isolated Pairs. \citet{Rampazzo1992} estimated that between as much as 10-25\% of the pairs in any complete
%(non-hierarchical) sample will be of the mixed morphology type. At the beginning of 1990s, studies about this kind of pairs were addressed to ascertain possible enhancement of the star formation activity, with
%respect to non interacting sample, via mid and far infrared observations,  at that time often hampered by a low resolution   \citep[see e.g.][] {Xu1991, Surace1993}. Mixed morphology pairs have
%been thought as the cleanest systems where to verify possible mass transfer between the gas rich and the gas poor member, typically the early-type companion. Several candidate of mixed morphology pairs with star
%formation and AGN activity fueled by gas transfer between components have been indicated \citep[see e.g.][]{deMello1995,Rampazzo1995,Domingue2003}.

In the NGC~454 system, there is no evidence of a velocity difference between the two members. The pair is furthermore strongly isolated as discussed in Appendix~\ref{A1:enviroment}.
Our observations do not provide direct evidence of gas re-fueling of NGC 454~E on the part of NGC~454~W. We found, however, traces of ionized gas beyond the nucleus
as shown in Figure~\ref{figure5}. We detect broad emission line in the NGC~454~E nucleus but the small Free Spectral range prevents us to disentangle 
the presence of multiple components which may provide us information about possible gas infall  
\citep[see e.g.][and references therein]{Rampazzo2006,Font2011,Zaragoza-Cardiel2013}. 
However  velocity gradient of about 130 km~s$^{-1}$ have been revealed in the central 4\arcsec.
{\it Swift}-{\tt UVOT} observations of NGC~454~E suggest that the galaxy is an S0 since a disk emerge in all the {\tt UVOT} bands when a S\'ersic law is fitted to the luminosity profile. Both the 
($B-V$) and ($M2-V$) color profiles become bluer with the galacto-centric distance supporting the presence of a disk \citep[see also][]{Rampazzo2017}. The disk is itself
strongly perturbed by the interaction as can deduced by the distortion in the NGC~454~T area (Figure~\ref{figure1}).

Both \citet{Johansson1988} and \citet{Stiavelli1998} speculated
whether the morphology of NGC~454~W pair member
was a spiral or an irregular galaxy, before the interaction.  
The galaxy, strongly star forming,
dominates in the NUV images with respect to NGC~454~E.
%In contrast, at longer wavelength NGC454~E dominates and
%NGC~454~W does not show unambiguously a bulge structure.
There is one evidence coming out from our velocity map: no rotation pattern
are revealed, even in NGC~454~W1 complex. The velocity difference
between the complexes  reaches $\approx$140 km~s$^{-1}$ (Figure~\ref{figure4})
if we include the NGC~454~SE an SW complexes, and it is about 60  km~s$^{-1}$
considering only NGC~454~W1-W6. 
To summarize, if NGC~454~W was a former spiral galaxy it appears completely
distorted by the encounter and this latter is not at an early phase.  
%However, this does not completely explain the lack of a bulge. Likely a much
%later type, a bulge less spiral or an irregular galaxy, together with the
%interaction could account both for the  lack of the nucleus and of a rotation pattern.
In the next section, we will investigate the  formation and evolution of this pair highligthening  
a possible interaction which matches the global properties of this pair, i.e.  its total magnitude, morphology and multi-wavelength SED.

%Concerning the evolution of the NGC~454~SW-SE complexes, another
%interesting point is the economy of the encounter. The two
%stellar complexes have a velocity difference $\Delta$V$\approx$100-120 km~s$^{-1}$
%with W1-W6 stellar complexes forming the body of NGC~454~W.

NGC~454~SW, SE have a projected separation of $\approx$37\farcs19 (8.7 kpc) 
and $\approx$39\farcs86 (9.4 kpc) from the center of W1, i.e. they occupy
a very peripheral position with respect to the bulk of the galaxy.
Figure~\ref{figure1} and Figure~\ref{figure2} both show that 
there is a very faint connection with the rest of the galaxy. 
%At the beginning of 1990s the research about tidal dwarf was quite fresh as well as the
%question about the nature and the evolution of extragalactic globular
%clusters.   The first starts to be posed theoretically by a paper by
%\citet{Barnes1992}. The second was reviewed by \citet{Harris91}
%\citep[see also][]{Harris01}.
%In this context, \citet{Johansson1988} speculated that the 
% NGC~454~SW and SE stellar complexes are possibly newly formed
% globular cluster. He calculated that their masses are of the
 %order of 3$\times$10$^{6}$ M$_\odot$ and their magnitudes,
 %after 10$^9$ years, will be comparable to that of  M31 globular clusters. 
 Our  kinematical study suggests that the two complexes
 do not show a rotation pattern. So NGC~454~SW and SE differ
 from tidal dwarf candidates as described in  \citet{Lelli2015}. Figure~\ref{figure6}
 shows that H$\alpha$ emission lines, detected  also in between the
 SW and SE stellar complexes,  are not composed of multiple
 components. 
 According to the scheme proposed by \citet{Moiseev2012} 
 the ($I-\sigma$) plots (Figure~\ref{figure7})  suggests that
 the ionized gas in SW and SE stellar complexes have
 the characteristic  of dense H\,{\textsc{ii}}
 regions surrounded by low-density gas with considerable 
 turbulent motions (see the scheme in their Figure 6)
 not dissimilar from W2-W6 complexes. 

To summarize both the {\it Swift}-{\tt UVOT} NUV observations and
 the diffuse H$\alpha$ emission indicate that NGC~454~W, NGC~454~SW
 and SE complexes are strongly star forming regions.
The anatomy of these complexes made using ($I-\sigma$), ($I-V_r$) and
 ($V_r-\sigma$) diagnostic diagrams indicates that 
 H\,{\textsc{ii}}  shells of different ages are present as well as zones of gas turbulence
 as expected by the interplay of star formation and SNae explosion in the IGM.   
%\citet{Combes1994} suggest that one of the main phenomena triggering star
%formation in pairs is the enhancement of the total amount of molecular gas.
%Gravitational torques induce gas  inflows that, enhancing the gas mass available,
%propel star formation. Recently \citet{Zaragoza-Cardiel2017} investigated  the
%kinematics of the ionized and molecular gas in three Arp interacting systems,
%namely Arp 186, Arp 236 and Arp 298. They found that the surface densities
%of the star formation rate and the molecular gas are significantly higher in these galaxies
%than in non-interacting galaxies and in the Milky Way itself, being closer to high redshift galaxies. 
Although our observations did not reveal direct evidence
of gas infall on the center of NGC~454~E there is  signatures of recent star formation,
in addition to the non thermal Seyfert 2 emission. \citet{Mendoza-Castrejon2015}
reported  the presence in {\it Spitzer}-IRS spectra of polycyclic aromatic hydrocarbons
(PAHs) which are connected to recent star formation episodes \citep[see eg.][and references 
therein]{Vega2010,Rampazzo2013}.

%%%%%%%%%% FIGURE 11 - SPRECTRAL ENERGY DISTRIBUTION %%%%%%%%%%%
  \begin{figure}
  \includegraphics[width=8.5 cm, angle=0]{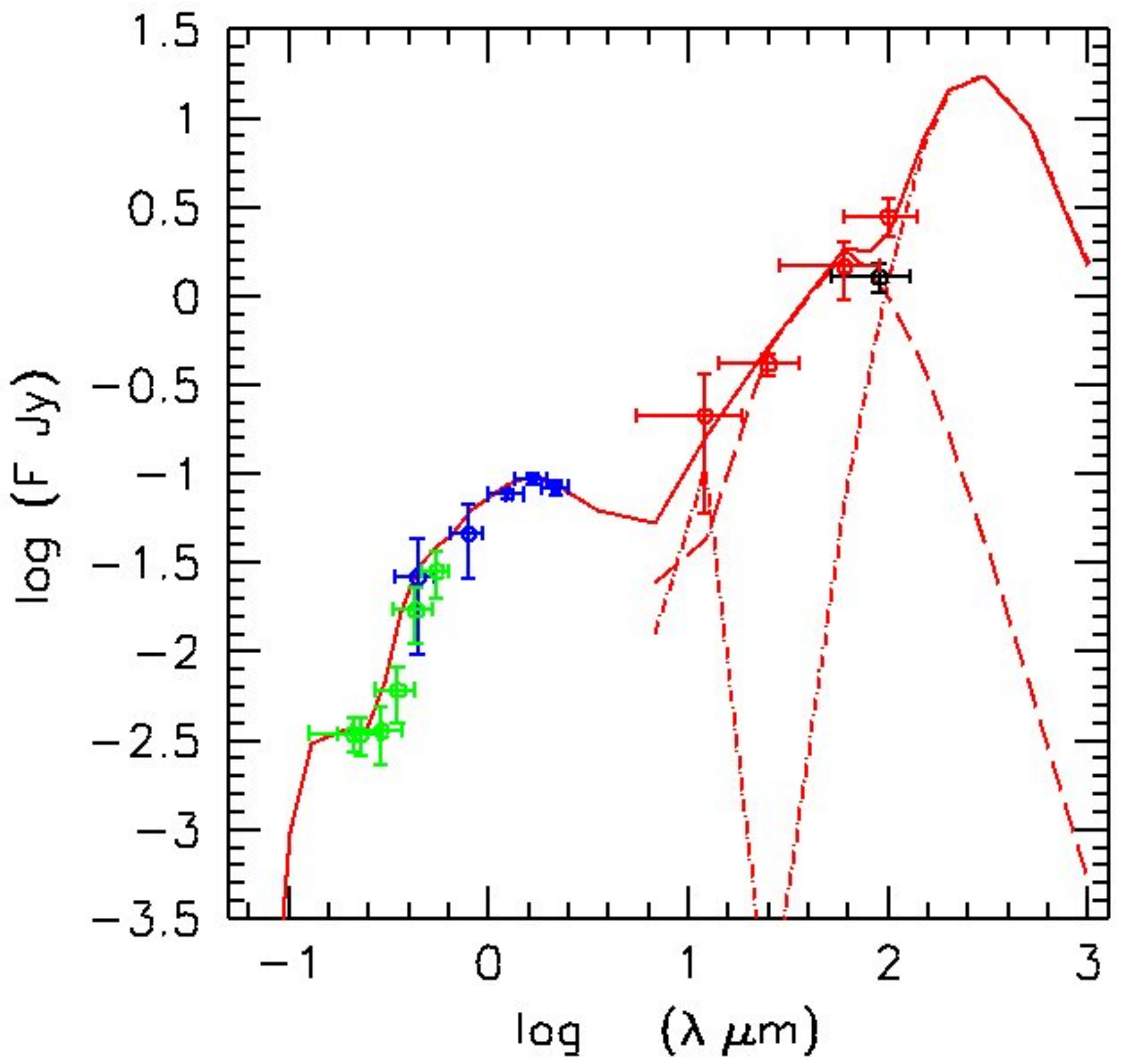}
  \caption{Observed spectral energy distribution(SED) of the whole NGC 454 system.
  The contribution of both the dust components to the FIR SED is also shown: 
  dot-dashed and the long-dashed lines are the warm and the cold dust componenr, respectively  (see text). 
  The solid line represents the resulting SED. Green symbols represent
 our {\it Swift}-{\tt UVOT} observations; blue symbols B, R, IRAS and 2MASS J, H, K 
 measures. Red symbols are {\tt IRAS} data while the black symbol is the 
 {\tt AKARI/FIS} detection.}
  \label{figure11}
  \end{figure}
%%%%%%%%%%%% END FIGURE 11  %%%%%%%%%%%%%%%

%%%%%%%%% FIGURE 12 - IMAGE SIMULATION %%%%%%%%%%%%%%%%
  \begin{figure}
  \includegraphics[width=8.3 cm, angle=0]{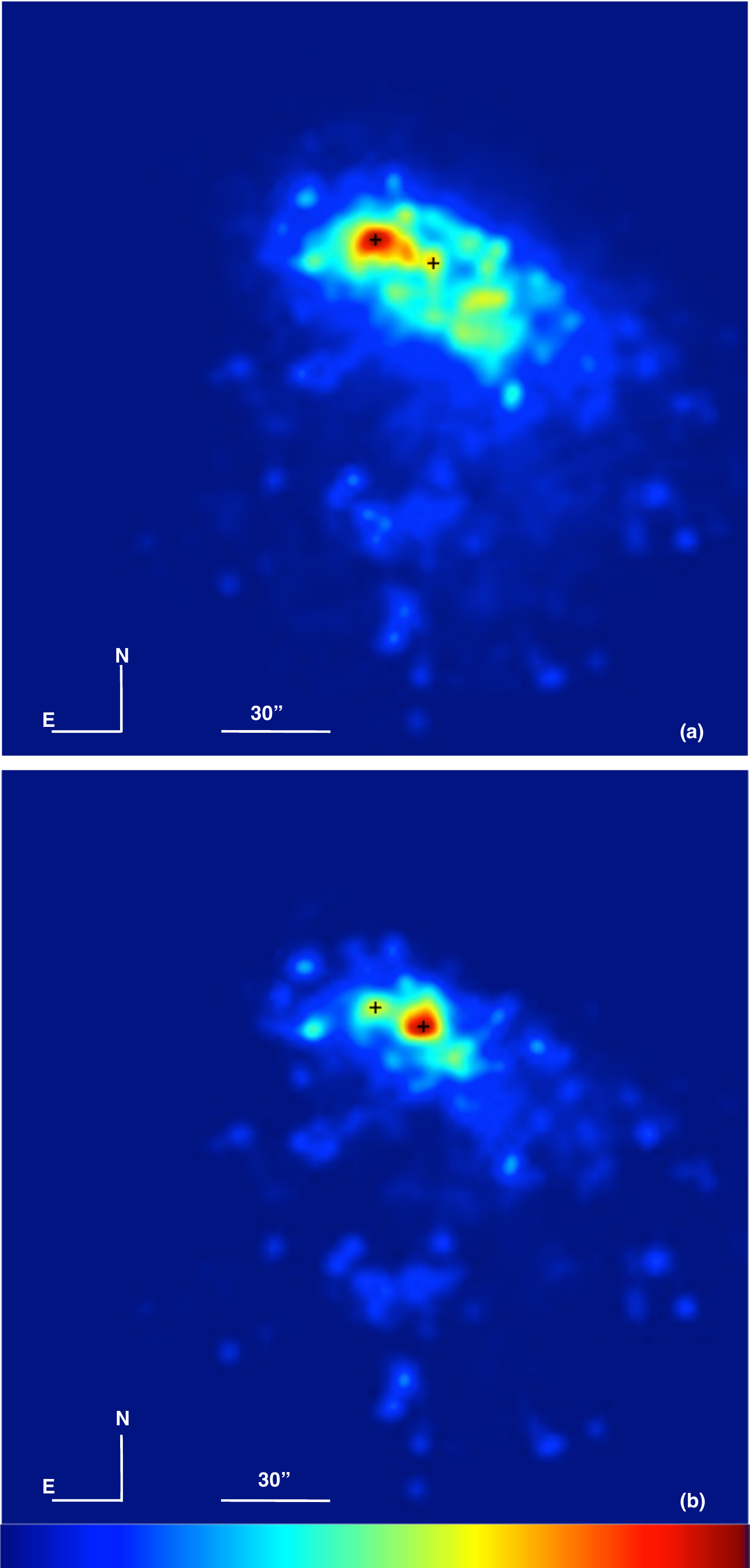}
  \caption{V-band  (a)  and UV (M2-band, (b))  xz projection of our
  simulation at the best-fit age; maps are normalized to the total
  flux within the box, and account for dust attenuation with the same
  recipes used to provide  the SED in Figure~\ref{figure11}. 
  The  scale is as in Figure~\ref {figure2}, with  the density contrast
  being equal to 100. Crosses  emphasise  the nuclei of  the merging
  galaxies, corresponding to  E and Irr in Figure~\ref{figure2} and
  N-E directions are given to guide the comparison.}
  \label{figure12}
  \end{figure}
%%%%%%%%%%%%%%%%%%%%%%%%%%%%%% END FIGURE 12  %%%%%%%%%%%

\subsection{Possible evolutionary scenario}
 \label{subsec:simulation}
 
%------------ table 5 input SPH---------------------
\begin{table*}
%\scriptsize
\center
\caption{Input parameters  of SPH-CPI simulation of NGC 454}
\begin{tabular}{llccccccccc}
\hline
 N$_{part}$&a &p/a& r$_1$ & r$_2$ & v$_1$ & v$_2$ & M$_T$ & f$_{gas}$\\
   &[kpc]&  & [kpc]&   [kpc] & [km/s] & [km/s]  & [$10^{10}\,M_\odot$] & \\
\hline\hline
 60000 &1014 & 1/3 & 777  & 777 &57&57& 400& 0.1   \\
\hline
\end{tabular}

\label{table5}

\footnotesize{Columns are as follows: (1) total number of initial (t=0) particles; (2) length of  the semi-major axis of the halo; (3) peri-centric separation of the halos in units of the semi-major axis; (4)  and (5)  distances of the halo centres of mass from the centre of mass of the total system, (6) and (7) velocity moduli of the halo centres in the same frame; (8) total mass of the simulation; (9) initial gas fraction of the halos.}
\end{table*}
%-------end table 5 input SPH---------------------

We investigate the evolution of the NGC 454 system using smooth particle
hydrodynamical (SPH)  simulation with chemo-photometric implementation
\citep[and references therein]{Mazzei2014}. 
Simulations have been carried out with different total mass (for each system  from 10$^{13}$\,M$_{\odot}$ to 10$^{10}$\,M$_{\odot}$), mass ratios (1:1 - 10:1),  gas fraction (0.1 - 0.01) and particle number (initial total number from 40000 to 220000).
All our  simulations of galaxy formation and evolution start from collapsing triaxial systems (with triaxiality ratio, $\tau$=0.84 \citep{Mazzei2014}, composed of dark matter (DM) and gas and include self-gravity of gas, stars and DM, radiative cooling, hydrodynamical pressure, shock heating, viscosity, star formation, feedback from evolving stars and type II supernovae, and chemical enrichment as in \citet{Mazzei03}.
We carried out different simulations varying the orbital initial conditions in order to have, for the ideal Keplerian orbit of two points of given masses, the first peri-centre separation, p, equal to the initial length of the major axis of the more massive triaxial halo down to 1/10 of the same axis. For each peri-centre separation we changed the eccentricity in order to have hyperbolic orbits of different energy. The spins of the systems are generally parallel each other and perpendicular to the orbital plane, so we studied direct encounters. Some cases with misaligned spins have been also analysed in order to investigate the effects of the system initial rotation on the results.
From our grid of simulations
we single out  the one which simultaneously (i.e.,  at the same snapshot) 
accounts for the following observational constrains providing  the
best-fit of the global properties of  NGC~454: i)  total absolute B-band
magnitude within the range allowed by observations (see below); ii) the 
 predicted spectral energy distribution (SED hereafter)  in agreement
with the observed  one; iii)  morphology like the observed  one in the
same bands and with the same spatial scale (arcsec/kpc). The results we
present are predictions of the simulation which best reproduces all the
previous observational constrains at the same snapshot. This snapshot
sets the age of the galaxy.\\
\indent
To obtain the SED of the whole  system extended over the widest
wavelength range,  we add to our UV and optical {\it Swift}-{\tt UVOT}
total fluxes  (green points in Figure~\ref{figure11})   the   B, R and
IRAS data (red points in Figure~\ref{figure11}) in {\tt NED}, and the
J, H, K   total fluxes, derived from 2MASS  archive images, which
perfectly agree with  J, H and K values reported by \citet{Tully2015}. 
All these data are corrected for galactic extinction as reported in
\S2.2 (Table 3) and \S6.  The black point in Figure~\ref{figure11}  is
{\tt AKARI/FIS}  detection. The solid line (red)  in
Figure~\ref{figure11} highlights the predicted SED. \\ 
\indent
The simulation which provides this fit corresponds to a major merger between 
two halos, initially of dark matter and gas, of  equal mass and gas fraction
(0.1),  with perpendicular  spins and total mass  4$\times$10$^{12}$\,M$_{\odot}$. 
Their mass centres are initially
1.4\,Mpc away each other and move at relative velocity of 120\,kms~$^{-1}$.
 Table~\ref{table5} reports the input parameters of the 
SPH-CPI simulation best fitting the global properties of the system.
The age of the system is 12.4\,Gyrs at the best-fit. The far-IR SED
accounts for a B-band attenuation of 0.85 mag so that the absolute
B-band magnitude of the best-fit snapshot  is -21.0 mag. This is the
value to be compared with that derived from the distance in Table
\ref{table1} accounting for an error  of $\pm$3.5\,Mpc  and a total
B-band magnitude  of 12.66$\pm$0.21 mag (from Table \ref{table3}) , that
is M$_B$=-20.64$\pm$0.41 mag.

Our fit of the far-IR emission implies a warm dust component, heated by
the UV radiation of H\,{\textsc{ii}}  regions, and a  cold component heated
by the general radiation field, both including PAH molecules as
described in \citet{Mazzei1992, Mazzei1994}, and
\citet{MazzeideZotti1994}, with a relative contribution, r$_{w/c}$=0.5
which means that  50\% of the far-IR emission is due to warm dust
emission. The cutoff radius of the cold dust distribution in
Figure~\ref{figure11} is 100\,r$_c$, r$_c$ being the core radius 
\citep{Mazzei1994}.

The shape of the far-IR SED suggests the presence of  a large amount of
dust:  the ratio between the the far-IR  luminosity  and the observed luminosity
in the UV to near-IR spectral range is 2.5.

Figure \ref{figure12} shows the morphology of our simulation at the
best-fit age to be compared with that in Figure~\ref{figure2}. We point
out that  the West component dominates the UV morphology and the Est
component the optical one, as observed. The simulation shows that these
systems will merge within 0.2\,Gyr.

Therefore, our approach points towards  a picture where E+S pairs can be
understood in terms of 1:1 encounters giving rise to a merger in less
than 1\,Gyr. Of course, this framework deserves further investigation,
that is beyond the scope of this work.

\section{Summary and Conclusions} 
\label{sec:summary}

We used SAM-FP observations  at SOAR
and {\it Swift}-{\tt UVOT} archival images to investigate the kinematical
and photometric properties of the NGC~454 interacting/merging system. 

According to the definition in \citet{Stiavelli1998}, we subdivided the system in
NGC~454~E, the early-type member, NGC~454~T, the perturbed area to the North of
NGC~454~W, the late-type member, and the two NGC~454~SW and SE complexes
South of the late-type. Further subdivision of the NGC~454~W member into W1-W6 
have been used to detail single H$\alpha$ complexes revealed by the monochromatic map
\citep[see also][]{Johansson1988}.

We found the following results:

The H$\alpha$ map shows that the emission is mostly detected in the NGC~454~W system and in the NGC~454SW and NGC~454~SE complexes. 
A H$\alpha$ broad emission is revealed in the center of NGC~454~E, with a velocity gradient of 130 km~s$^{-1}$ across 4\arcsec.

The radial velocity map does not have a rotation pattern neither in the
W1-W6 complexes in NGC~454~W, nor in the two SW and SE complexes. W6
shows a velocity gradient of 45 km~s$^{-1}$.

The velocity dispersion map shows that most of the W3-W6, SW and SE
complexes have a velocity dispersion in the range 20-25 km~s$^{-1}$. The
highest velocity dispersion, 68 km~s$^{-1}$ and the lowest, 15
km~s$^{-1}$, are measured in the W1 and W2 complexes, respectively.

We use ($I-\sigma$) ($I-V$) ($\sigma-V$) \citep[see eg.][]{Bordalo2009}
diagnostic diagrams to study the kinematics of the W1-W6 complexes in
NGC~454~W and the SW and SE complexes. Diagnostic diagrams show
that all regions, except W1 and SW1, have a weak-moderate correlation between
the radial velocity and the dispersion interpreted as systematic motions toward or away from the observer. 
These diagrams confirm that W1 has high supersonic velocity dispersion
and a closer analysis could separate two populations, one in the center
with a low dispersion and a second, around it with a higher $\sigma$.
According to \citet{Moiseev2012}, W1 could show a Giant HII Region in the center and a turbulent low-density gas cloud in its outskirt.
This picture can
be discussed further if we take into account the several large profiles
in W1 show multiple components. If we only take into account the main
component (the brightest), the situation is reversed, with a broader
line in the center and narrowed ones around. This can be interpreted as
a expanding wind blow bubble according to \citet{Lagrois2009}.

Based on our SPH simulation with chemo-photometric implementation, the
global properties of the system, 12.4 Gyr old, are compatible with an
encounter between two halos of equal mass and perpendicular spin. They
will merge within 0.2 Gyr. The SED suggests a large FIR emission 2.5
times that in the NUV-NIR range.

The case of NGC 454 system suggests  that a class of mixed pairs form
via the encounter/merging of similar mass, evolving halos. The different
morphologies, emphasized by multi-$\lambda$  observations, mark a late 
phase of the merging process.

\section*{Acknowledgements}

Authors would like to warmly thanks A. Tokovinin, C. Mendes de Oliveira and B. Quint to their participation during the run and 
without which the commissioning of the instrument and the run would not have been possible. Authors would like to thank the anonymous referee for its useful corrections. HP  thanks SOAR staff. 
HP thanks Aix Marseille Universit\'e for its financial support during his visit Laboratoire d'Astrophysique de Marseille in April-July 2017.
Paola Mazzei and Roberto Rampazzo acknowledge support from INAF through grants 
PRIN-2014-14 'Star formation and evolution in galactic nuclei' and
PRIN-SKA 2016 'Empowering SKA as a probe of galaxy evolution with HI'. We
acknowledge the usage of the {\tt HyperLeda} database
(http://leda.univ-lyon1.fr). Part of this work is based on archival
data, software or online services provided by the ASI SCIENCE DATA
CENTER (ASDC). We also acknowledge the usage of the Nasa Extragalactic Database (http://ned.ipac.caltech.edu/) and R free software.

%%%%%%%%%%%%%%%%%%%%%%%%%%%%%%%%%%%%%%%%%%%%%%%%%%
%%%%%%%%%%%%%%%%%%%% REFERENCES %%%%%%%%%%%%%%%%%%
%%%%%%%%%%%%%%%%%%%%%%%%%%%%%%%%%%%%%%%%%%%%%%%%%

% The best way to enter references is to use BibTeX:

%\bibliographystyle{mnras}
%\bibliography{example} % if your bibtex file is called example.bib

% Alternatively you could enter them by hand, like this:
% This method is tedious and prone to error if you have lots of references

%%%%%%%%%%%%%%%%%%%%%%%%%%%%%%%%%%%%%%%%%%%%%%%%%
%%%%%%%%%%%%%%%%% APPENDICES %%%%%%%%%%%%%%%%%%%%%
%%%%%%%%%%%%%%%%%%%%%%%%%%%%%%%%%%%%%%%%%%%%%%%%%

\appendix
%
%
%If you want to present additional material which would interrupt the flow of the main paper,
%it can be placed in an Appendix which appears after the list of references.
%
\section{The environment of the NGC~454 system}
\label{A1:enviroment}

 \begin{table*}
   % \begin{minipage}[t]{\columnwidth}
   \caption{Galaxies in 4$\times$4 Mpc the  NGC~454 system}
   \label{table-a1}
   \centering
   \renewcommand{\footnoterule}{} % to avoid a line before footnotes
    \begin{tabular}{llcccccccc}
    \hline \hline
   Ident.     &  Other Ident. &  RA (2000)  &  D (2000)   & T  &  V$_h$  & logd$_{25}$ & 
   log~r$_{25}$  &  PA 
   & 
   B$_T$ \\
                &                F      & [deg.]         &   [deg.]      &      & [km~s$^{-1}$] & 
                log[0.1\arcmin]  &  & [deg.] &  [mag] \\
   \hline 
ESO113-004 & AM 0058-580 & 1.00950& -57.74830 &  5.0 &3566$\pm$67& 0.75 & 0.19 
&105.5& 
15.04$\pm$0.20\\
ESO113-009 &AM 0102-573 & 1.08046 & -57.37742 &10.0 & 3648$\pm$6  & 0.99 &  0.54 
&165.4 & 
16.02$\pm$0.20\\
NGC~454~W & RR 023a &1.23945 & -55.40013 & -1.0 & 3626$\pm$2  & 1.22  & 0.00 & \dots 
&13.12$\pm$ 0.20\\
NGC~454~E & AM 0112-554; RR 023b & 1.24025 & -55.39714 & -2.0 & 3635$\pm$2   & 1.28 
&  
0.38 & 80.9 
&13.14$\pm$0.20\\
ESO113-044 & \dots & 1.37632 & -57.56919 & 10.0 & 3616$\pm$8  & 0.67  & 0.15 & 53.3 
&16.99$\pm$0.20\\
PGC005497 & AM 0126-525 &1.47367 & -52.63588  & 9.0 & 3450$\pm$42  & 0.92  & 0.34 & 
92.7 
&16.78$\pm$0.20\\
NGC0576 & AM 0126-515 & 1.48269 & -51.59871 & -1.1 & 3604$\pm$39 &  0.99 &  0.08 &  
\dots 
&14.42$\pm$0.19\\
ESO196-011 & \dots & 1.51217 & -51.14189 & 5.8 & 3634$\pm$6 &  1.25 &  0.42 & 12.6 
&14.47$\pm$0.20\\
  \hline
   \hline
    \end{tabular}
    
  % \end{minipage} 
  
For each galaxies the columns provide the following information: (1) the galaxy identification, (2.3) 
the (J2000)  galaxy coordinates, (4) the morphological type, (5) the heliocentric velocity, (6) the  log 
of the length the projected major axis of a galaxy at the isophotal level 25 mag~arcsec$^{-2}$ in the 
B-band, 
(7) the log of  the axis ratio (major axis/minor axis) of the isophote at 25 mag~arcsec$^{-2}$ in the  
B-band , (8) the position angle, and (9) the total 
apparent B-band magnitude. Data are from {\tt Hypercat}. Other identification are from {\tt NED}. 
   \end{table*}
%%%%%%%%%%%%%%%%%%%%%%%%%%%%%%%%%%%%%%%%%%%%%%%%%%%%%%%%%%%%%%%%%%%%%%%%%%%%%%%%%%%%%%%%%%%%%%
%%%%%%%%%%%%%%%%%%%%%%%%%%%%%%%%%%%%%%%%%%%%%%%%%%%%%%%%%%%%%%%%%%%%%%%%%%%%%%%%%%%%%%%%%%%%%%%%%%

The NGC 454 system is an isolated pair, RR 23, 
in the \citet{RR95} catalogue. Assuming the distance given in Table~\ref{table1}
we used  {\tt Hyperleda} to inspect a box of 4$\times$4 Mpc$^2$ for possible 
neighbors of this nearby system. 
Table~\ref{table-a1}, which includes the pair members,
provides the galaxy identification (col. 1 and col. 2), the right ascension
and declination (col.s 3 and 4), the morphological Type (col. 5),
the heliocentric velocity (col. 6), the major axis diameter $d_{25}$ (col. 7) and
the axial ratio $r_{25}$ (col. 8) at $\mu_B$=25 mag arcsec$^2$, the
position angle (col. 9) and the total B-band apparent magnitude (col. 10).

Most the galaxies in Table~\ref{table-a1} are listed in {\it  A Catalogue
of Southern Peculiar Galaxies and Associations} \citep{AM1987} where they
are recognized either to show peculiar fearures (AM 0058-580 compact galaxy 
with diametric jets; AM 0102-573 disrupted galaxy + 2 companions; AM 0126-525
ring or galaxy with loop; AM 0126-515 disturbed spiral)
or to be pair members (like our NGC 454 system i.e. AM112-554 I/A double + resolved knots)

\begin{figure}
 \includegraphics[width=8.5cm]{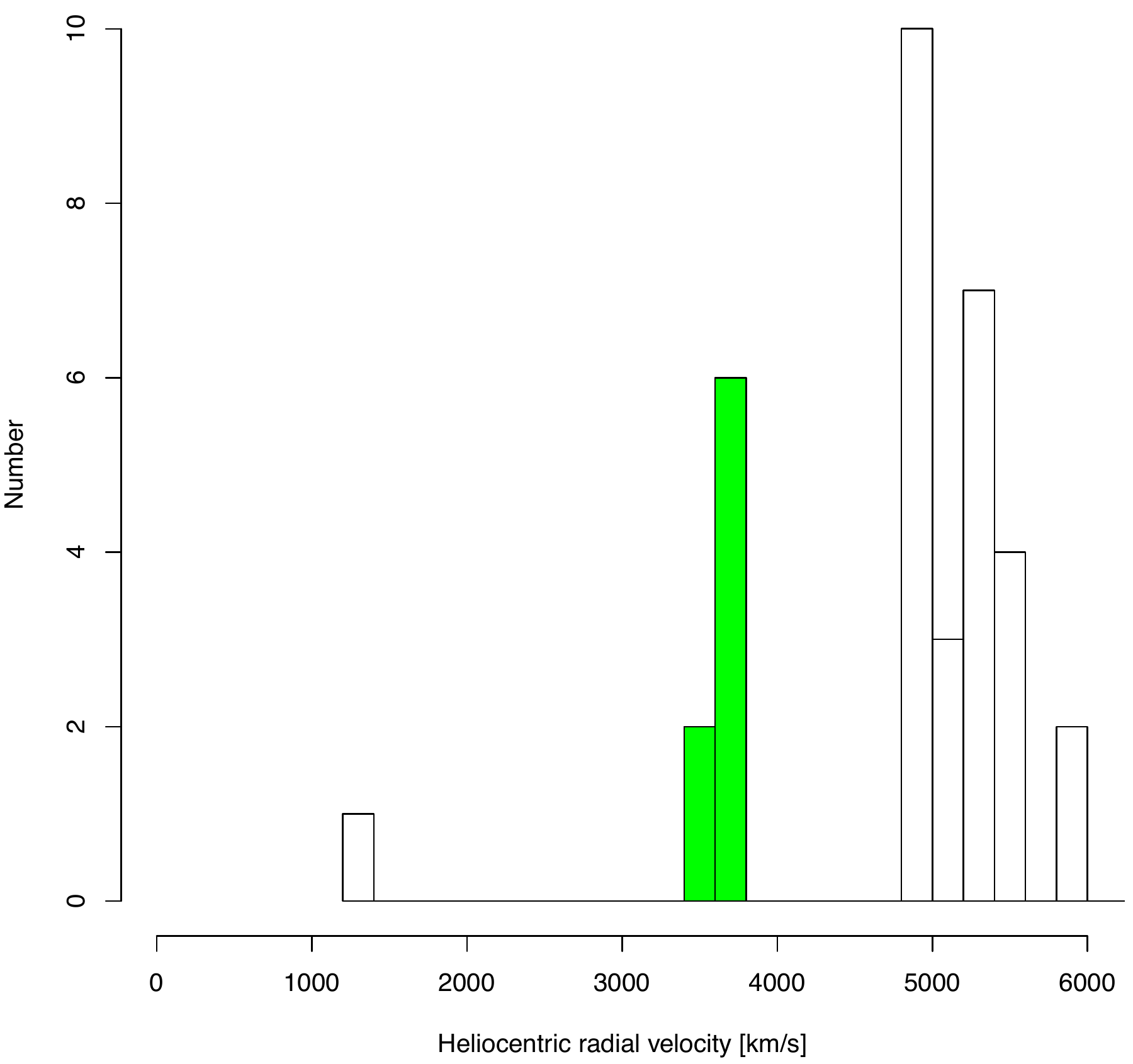}
 \includegraphics[width=8.5cm]{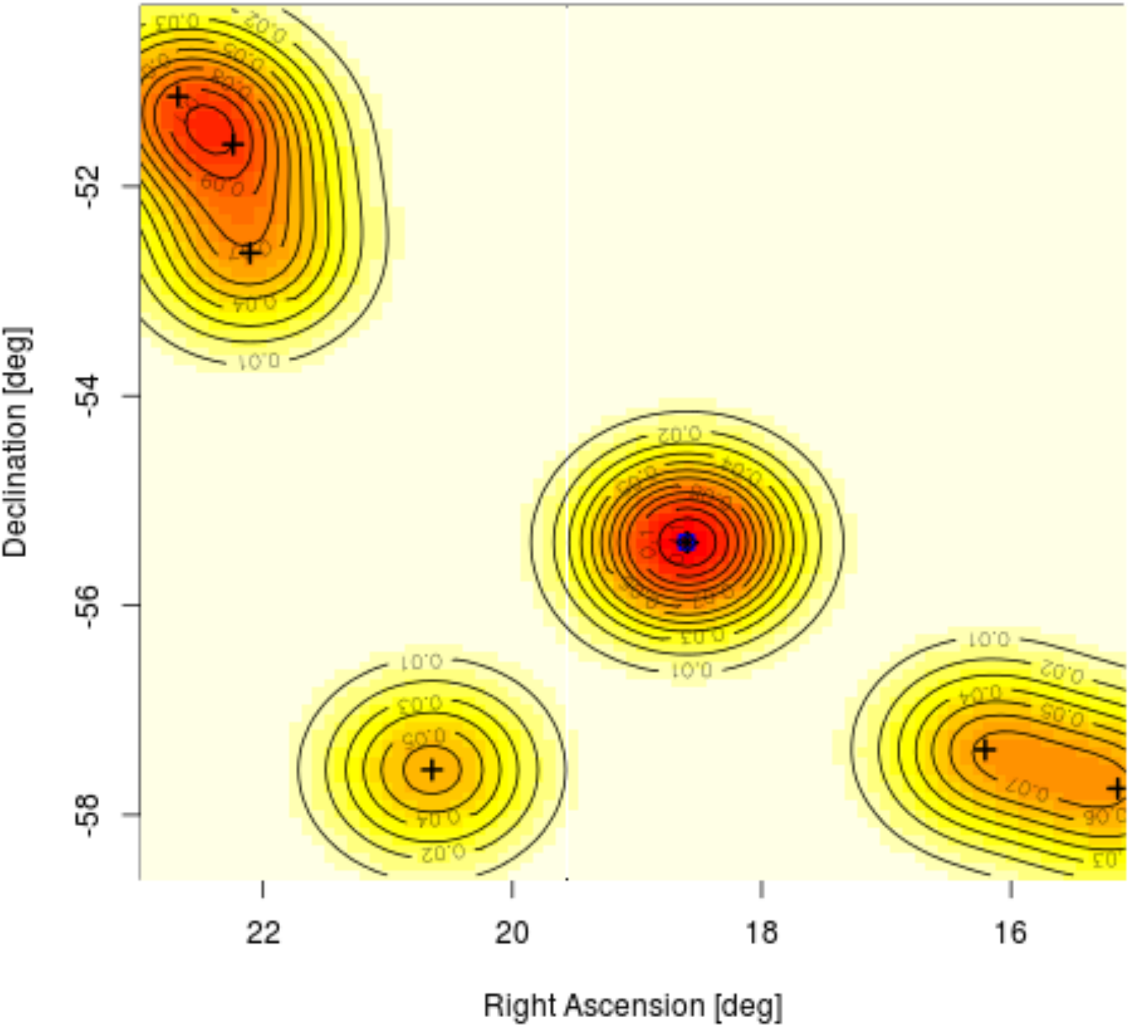}
 \caption{(top panel) Histogram of the heliocentric radial velocity of
 galaxies within a box of 4$\times$4 Mpc$^2$ in the range 1000 -- 6000
 km~s$^{-1}$. (bottom panel) Spatial distribution of galaxies in green
 in the top panel, i.e. the neighbors of the NGC~454 system. The map is
 normalized to the total density. Galaxies in the area are listed in
 Table~\ref{table-a1}. }
 \label{figure-a1}
 
 \end{figure}

Figure~\ref{figure-a1} shows the histogram of the recession velocity
distribution (1000 -- 6000 km~s$^{-1}$) in an area of 4$\times$4 Mpc$^2$
around NGC~454 and the spatial distribution in the same area of the
nearby galaxies, shown in green in the top panel.  The nearby neighbors are
disk galaxies either S0s or Spirals according to the classification of {\tt Hyperleda}.
The velocity dispersion of these galaxies is 65 km~s$^{-1}$.
This value has  to be compared with 327$^{+12}_{-2}$ km~s$^{-1}$ of NGC 5486, 
the third rich  galaxy association in the nearby Universe \citep{Marino2016} and with
92 $^{+3}_{-2}$ km~s$^{-11}$ of LGG 225 a very loose group discussed
in \citet{Mazzei2014}. This picture confirms that NGC~454 system is isolated and
located in a very poor environment \citep[see also][]{Tully2015}. 
 
%%%%%%%%%%%%%%%%%%%%%%%%%%%%%%%%%%%%%%%%%%%%%%%%%%%%%%%%%%%%%%%%%%%%%%%%%%%%%%%%%%%%%%%%%%
%%  -------------------  FIGURE A1 
%%%%%%%%%%%%%%%%%%%%%%%%%%%%%%%%%%%%%%%%%%%%%%%%%%%%%%%%%%%%%%%%%%%%%%%%%%%%%%%%%%%%%%%%%%

%%%%%%%%%%%%%%%%%%%%%%%%%%%%%%%%%%%%%%%%%%%%%%%%%%%%%%%%%%%%%%%%%%%%%%%%%%%%%%%%%%%%%%%%%%
%%                      END FIGURE A1
%%%%%%%%%%%%%%%%%%%%%%%%%%%%%%%%%%%%%%%%%%%%%%%%%%%%%%%%%%%%%%%%%%%%%%%%%%%%%%%%%%%%%%%%%%

% Don't change these lines
\bsp	% typesetting comment
\label{lastpage}
\end{document}